\documentclass[manuscript]{acmart}
\AtBeginDocument{%
  \providecommand\BibTeX{{%
    \normalfont B\kern-0.5em{\scshape i\kern-0.25em b}\kern-0.8em\TeX}}}

\copyrightyear{2022}
\acmYear{2022}
\setcopyright{acmcopyright}
\acmConference[FAccT '22]{2022 ACM Conference on
Fairness, Accountability, and Transparency}{June 21--24, 2022}{Seoul, Republic of
Korea}
\acmBooktitle{2022 ACM Conference on Fairness, Accountability, and Transparency
(FAccT '22), June 21--24, 2022, Seoul, Republic of Korea}
\acmPrice{15.00}
\acmDOI{10.1145/3531146.3533108}
\acmISBN{978-1-4503-9352-2/22/06}

\usepackage{color,soul,colortbl}
\usepackage{multirow}
\usepackage{wrapfig}

\begin{document}

\title{Interactive Model Cards: A Human-Centered Approach to Model Documentation}

\author{Anamaria Crisan}
\authornote{Both authors contributed equally to this research.}
\email{acrisan@tableau.com}
\orcid{0000-0003-3445-3414}
\author{Margaret Drouhard}
\authornotemark[1]
\email{mar.drouhard@tableau.com}
\affiliation{%
  \institution{Tableau Research \& User Research}
  \streetaddress{1621 N 34th St.}
  \city{Seattle}
  \state{Washington}
  \country{USA}
  \postcode{98103}
}

\author{Jesse Vig}
\email{jvig@salesforce.com }
\affiliation{%
  \institution{Salesforce Research}
  \streetaddress{575 High St }
  \city{Palo Alto}
  \state{California}
  \country{USA}
  \postcode{94301}
}

\author{Nazneen Rajani}
\email{nazneen@hf.co}
\affiliation{%
  \institution{Hugging Face}
  \city{Palo Alto}
  \country{USA}
}

\renewcommand{\shortauthors}{Crisan and Drouhard, et al.}


\begin{abstract}
Deep learning models for natural language processing (NLP) are increasingly adopted and deployed by analysts without formal training in NLP or machine learning (ML). However, the documentation 
intended to convey the model's details and appropriate use 
is tailored primarily to individuals with ML or NLP expertise. To address this gap, we conduct a design inquiry into \textit{interactive model cards}, which augment traditionally static model cards with affordances for exploring model documentation and interacting with the models themselves. Our investigation consists of an initial conceptual study with experts in ML, NLP, and AI Ethics, followed by a separate evaluative study with non-expert analysts who use ML models in their work. Using a semi-structured interview format coupled with a think-aloud protocol, we collected feedback from a total of 30 participants who engaged with different versions of standard and interactive model cards. Through a thematic analysis of the collected data, 
we identified several conceptual dimensions that summarize the strengths and limitations of standard and interactive model cards, including: stakeholders; design; guidance; understandability \& interpretability; sensemaking \& skepticism; and trust \& safety.  
Our findings demonstrate the importance of carefully considered design and interactivity for orienting and supporting non-expert analysts using deep learning models,
along with a need for consideration of
broader sociotechnical contexts and organizational dynamics. We have also identified design elements, such as language, visual cues, and warnings, among others, that support interactivity and make non-interactive content accessible. We summarize our findings as design guidelines and discuss their implications for a human-centered approach towards AI/ML documentation.
\end{abstract}


\vspace{-2mm}


\maketitle


\section{Introduction}
Open source development has made it easier to share and deploy complex models, including large language models. This ease-of-use has lowered barriers to non-expert analysts~\cite{yang2018grounding} who 
do not have
formal training in machine learning (ML), data science (DS), or linguistics.
To accommodate a spectrum of ML expertise,  Mitchell et al.~\cite{mitchell10.1145/3287560.3287596} have proposed model cards as a means of providing consistent summaries of model details and their potential for misuse and harm. The format they propose is text-based and concise, making it both broadly accessible and applicable across model types.
However, these model cards
and
similar forms of 
documentation rely on the developer to accurately and clearly report on the model and its performance. Often, this process is labor-intensive and many important details, such as unintended uses or disaggregated model performance, are omitted. 
Both experts and non-experts
who want to interrogate the model further must do so by implementing their own analysis. Not only does this mode of interrogation leave many non-experts underserved, it also exacerbates the potential for harm once these models are deployed~\cite{Crisan:fits_and_starts:2021,bender2021dangers,Benjamin2019AssessingRA,buolamwini:gender_shades:2018}. 

Prior work has shown that non-expert analysts benefit from interacting with machine learning models and their data~\cite{Amershi:Power:2014,dudley:IML:2018,scaha:HCD:2017}. Recently, others have experimented with adding interactive elements to model cards. Both HuggingFace~\cite{wolf:huggingfaces:2020} and Google Cloud platforms introduced interactive modalities for interrogating the model’s performance through customizable examples of the model’s inputs. Robustness Gym reports~\cite{goel2021robustness} 
seek
to overcome the development burden of model cards by allowing end-users to interactively create new slices of the data to interrogate a model's performance. Complementary to model cards are ‘explainables’, which are generally bespoke and interactive~\cite{hohman:explainable:2020}. While these early explorations are promising, they do not explore how much or what kind of interactivity is beneficial. 

In this work, we use the concept of a model card to scaffold a design inquiry into alternative and expanded forms of model documentation, focusing in particular on the needs of non-experts.  We propose a novel concept for an \textit{interactive model card} (IMC) that: (1) lowers barriers to accessing key information about model behavior; (2) supports deeper interrogation of models; and (3) surfaces some attendant risks and limitations without additional work on the part of model developers. We 
built
this idea out through an initial conceptual study with experts in ML, Ethics, and NLP to co-create a set of design guidelines for a functional IMC prototype. 
In this first study, we drew on experts' experiential knowledge of the wider implications of ML and NLP models, along with applications of these models within organizations, to 
\textit{better understand implications of design choices in model documentation and inform our development of an IMC.} 
We conducted a subsequent study with non-expert analysts to 
\textit{deepen our understanding of documentation needs and evaluate our IMC design.} 
Across both studies, we apply a human-centered lens to examine the affordances, opportunities, and limitations for standard model cards (SMCs)~\cite{mitchell10.1145/3287560.3287596}, Robustness Gym reports (RGRs)~\cite{goel2021robustness}, and IMCs.

\textit{\textbf{Through this research, we provide the following three contributions:}}
1) Design guidelines for interactive model cards and model documentation broadly;
2) A set of conceptual dimensions for evaluating model documentation with respect to AI/ML workers of 
various backgrounds; and
3) A functional prototype for an interactive model card that can be adapted for further inquiries and usage scenarios. We make artifacts of our research process available online\footnote{\url{https://osf.io/9d83t}}.

\section{Background and Related Work}


The \textbf{\textit{potential for both broad and deep harm}} that AI/ML systems can pose to vulnerable people and ecosystems, 
especially when impacted people do not have the opportunity to interrogate and contest decisions driven by AI/ML(e.g.,~\cite{barocas2016big,whittaker2018ai,hoffmann2019fairness,hamidi2018gender,hanna2020towards}). 
Large language models are especially predisposed to amplify harms,
since they are being deployed widely enough to result in homogenization while their inner workings and limitations are poorly understood~\cite{bommasani2021opportunities,bender2021dangers}. 
Notably, even systems that seek to identify and mitigate racism, harassment and hate speech
may compound harms to vulnerable people (e.g.,~\cite{benthall2019racial,corbett2018measure,mozafari2020hate}).
In an effort to mitigate the risks for harm in AI/ML systems, researchers and developers have proposed strategies to make systems more \textbf{\textit{explainable, transparent,} and \textit{contestable}} (e.g.,~\cite{guidotti2018survey,yang2021intellige,madsen2021post,arrieta2020explainable,kaur2020interpreting,mittelstadt10.1145/3287560.3287574,molnar2020interpretable}). 
Our study design (Sections~\ref{study_one:general_procedures} and~\ref{sec:study2_procedures}) has been informed by prior work related to measuring and improving model interpretability~\cite{poursabzi2021manipulating,mittelstadt10.1145/3287560.3287574}, and proposals for contrastive explanations~\cite{mittelstadt10.1145/3287560.3287574} are particularly aligned with our approach to interactivity and seeding the IMC with examples (as described in Section~\ref{study_two:functional-prototype}). Our approach to interactivity has also been informed by Amershi~\textit{et al.}'s guidelines for human-AI interaction~\cite{amershiGuidelines10.1145/3290605.3300233},
as well as by other guidelines and taxonomies for human-centered AI~\cite{arrieta2020explainable,shneiderman2020bridging} and proposals for enabling human contestation of AI-driven decisions~\cite{lyons10.1145/3449180,hirsch2017designing}. 

Prior work has also addressed dimensions of model behavior and documentation that impact \textbf{\textit{trust}} and \textbf{\textit{skepticism}}. 
Many studies have found that cognitive biases (particularly anchoring bias and automation bias more broadly) can lead users to misunderstand model behavior and place unwarranted trust in AI systems~\cite{kaur2020interpreting,Nourani:anchoring:2021,poursabzi2021manipulating,ehsan2021explainability}.
We introduce the concept of \textit{productive skepticism} (Section~\ref{sec:discussion}), and we argue that stimulating productive skepticism--along with providing modalities for interaction and sensemaking--could be an effective strategy to mitigate the tendency to over-trust. Rather, opportunities to engage with productive skepticism can support stakeholders in calibrating appropriate trust in a model's behavior for a particular context and use case.
Our approach 
aligns with calls for new strategies of calibrating user trust in AI~\cite{zhang10.1145/3351095.3372852} and promoting reflection~\cite{ehsan2021explainability}, and our approach to guidance and warnings (discussed in Section~\ref{study_one:design_guide}) encompasses the requirements for extrinsic trust articulated by Jacovi~\textit{et al.}~\cite{jacovi2021formalizing}.

Various forms of \textbf{\textit{model documentation}} have been proposed to make AI/ML systems more transparent and help users establish trust.
\textit{Model cards} (SMCs)~\cite{mitchell10.1145/3287560.3287596}  
have been foundational in standardizing documentation for AI/ML model performance and characteristics, along with 
risks
and unintended uses. 
Robustness Gym helps model developers report on model behavior 
through the generation of \textit{robustness reports} (RGRs) that summarize model performance on various data slices~\cite{goel2021robustness}. We incorporate SMCs and RGRs as core elements of the IMC design.
We also look to work on \textit{datasheets for datasets} to include more detailed information of the model's data~\cite{gebru2021datasheets}. The IMC design follows the SMC model for reporting on datasets, and our functional prototype (Section~\ref{study_two:functional-prototype}) also includes warnings related to age of the training dataset.
In evaluating the IMC,
we also draw on dimensions articulated 
for 
\textit{explainability fact sheets}~\cite{sokol10.1145/3351095.3372870}.

Model cards and similar standardized artifacts can play a role in processes for \textbf{\textit{governance}} and \textbf{\textit{accountability}} of AI/ML systems. As many have studies have reinforced, mechanisms for governance and accountability of technical systems are prerequisites for informed consent~\cite{giannopoulou2020algorithmic,carroll2020care,viljoen2021relational} and equity~\cite{dombrowski2016social,lee2019webuildai,krafft2021Action}.
Algorithmic audits are a key strategic approach toward governance and accountability.
Whether conducted through internal processes or by third-party actors, audits seek to evaluate AI models for issues of robustness, bias, and fairness.
Public audits have been shown to be effective at reducing bias in targeted companies (as compared to other companies) ~\cite{raji2019actionable}.
Auditing should be a structured, ongoing process that is designed from a system perspective~\cite{shneiderman2020bridging,kingsley2020auditing}. However, many challenges remain from conceptual, technical, economic, and organizational perspectives ~\cite{mokander2021ethics}.
In contrast to posthoc audits and evaluations, \textit{certificates} of robustness provide algorithmic guarantees on the performance of models under certain conditions ~\cite{zhang2021certified}. 
Human-in-the-loop auditing processes leverage everyday users to provide further protection once a system is deployed~\cite{shen10.1145/3479577}, and we heard from participants in both of our studies that they believed IMCs would be valuable for record-keeping and organizational alignment throughout deployment.

Recent research has surfaced implications for \textbf{\textit{practitioners' needs}} for trust and transparency.
Some have identified strategies practitioners use to build understanding 
of AI/ML models, including data- and model-centric patterns of exploration~\cite{mishra2021designing,sanchez10.1145/3449236} and investigation of data where a model performs particularly well and poorly~\cite{mishra2021designing,lucic10.1145/3351095.3372824,sanchez10.1145/3449236}.
Our considerations around trust and productive skepticism have been informed by prior work identifying domain expertise and predispositions for trust as risk factors in bringing models into practice~\cite{nourani2020role,Dietvorst:AlgorithmAP:2015,knijnenburg2011each,yang2018grounding,mishra2021designing}.
Furthermore, our support for exploration of sup-populations and sampling examples with low- and high-confidence scores address these needs in part.
Hong~\textit{et al.}~\cite{hong2020human} have also surfaced needs for integrated interpretability support with AI/ML tooling, which aligns with our findings.
In response to practitioners' needs to evaluate AI/ML models with respect to safety~\cite{wang2021ai} and ethical decision-making~\cite{holstein2019improving}, we have evaluated the IMC along these dimensions.
As many researchers have articulated, safe and effective deployment of AI/ML systems in practice requires a robust understanding of different stakeholders' needs and goals for explainability~\cite{bhatt10.1145/3351095.3375624,hong2020human}; support for traceability and auditing~\cite{wang2021ai}; and strategies to address concerns within existing organizational structures and dynamics~\cite{wang2021ai,holstein2019improving,bhatt10.1145/3351095.3375624,rakova10.1145/3449081,hong2020human}.
We found that our approach to guidance, shareability, and traceability begins to address these needs, and we surfaced additional relevant implications for design.

\vspace{-2mm}
\section{Study 1 : Concept Study with ML, AI, and NLP Experts}\label{sec:study_one}

Ahead of engaging with non-expert analysts, we leveraged the expertise of researchers and practitioners in  ML/AI development, AI Ethics, and Natural Language Processing (NLP), to co-create a set of design guidelines for IMCs.




\vspace{-2mm}
\subsection{Design Probes}\label{sec:studyone_probes}

\noindent\textbf{Design Scope for Interaction.} There are many ways that interactivity could be theoretically added to model cards. In this study, we have scoped the goals of interaction design to focus on disaggregated performance metrics, as well as understanding the model's training data and how it compares to the analyst's. 
We chose not to explore other sorts of interactivity, such as: comparing models (i.e., multiple sentiment analysis models), comparisons across multiple datasets, or training the model on the analyst's own data. We focus on disaggregated model performance and dataset understanding and comparison because we 
view these as key
tasks for understanding a single model that are undersupported with an SMC. Moreover, understanding the model's performance and data conforms to what we consider the spirit of the SMC, which is to inform the reader about a single model.


\begin{wrapfigure}{r}{0.5\textwidth}
    \centering
    \includegraphics[width=0.5\textwidth]{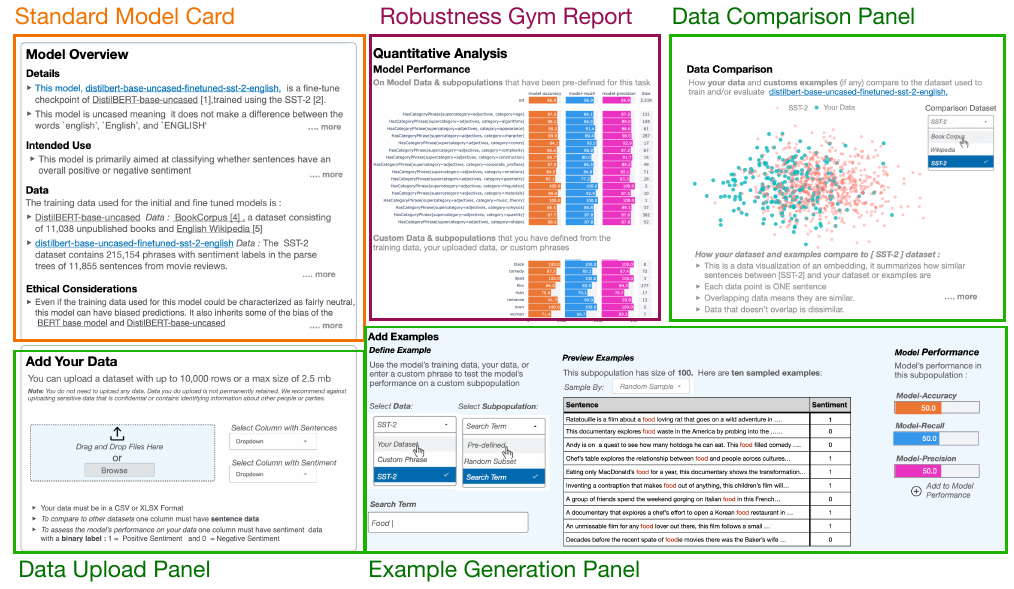}
    \caption{Concept of an Interactive Model Card. Larger images of the SMC, RGR, and IMC are available in the \href{https://osf.io/9d83t/?view_only=7359695185b04468a7ae4a2e189691c0}{online materials}}
    \Description{An illustration of the interactive model card conceptual probe that was used to in the first study with AI/ML experts. The probe is annotated with boxes in three colors : orange, purple, and green that show elements of the standard model card, robustness gym report, and the new elements of the interactive model card - respectively. The Standard model card sections contain abbreviated text from the SMC, which we describe in the main text. The robustness gym report shows a bar chart with three facets; each facet is a different model metric. The rows of the bar chart show different slices of the data that are automatically generated by Robustness Gym. The three new elements of the interactive model card as a follows. First, a data upload panel that sits underneath the standard model card and allows users to upload their own dataset. Next, a data comparison panel that is adjacent to the Robustness Gym report and shows a scatter plot to the dataset that the model was trained on. Below that is the Example Generation panel where the user can dynamically create new sentence or data set slices to evaluate the model on. }
    \label{fig:phaseI_concepts}
    \vspace{-5mm}
\end{wrapfigure}

\noindent\textbf{Reference Model Card: Task, Model, and Data}. We use a reference model card of DistilBERT~\cite{sanh:distilbert:2020} fine-tuned on the SST-2~\cite{socher:sst-2:2013} dataset for a sentiment analysis task; this reference mode is sourced from \href{https://huggingface.co/}{HuggingFace}~\cite{wolf:huggingfaces:2020}. We chose to focus on a sentiment analysis task because it is widely known and easily understood, while still being representative of NLP model development regimens for other tasks. Moreover, the DistilBERT SST-2 fine-tuned model presented a unique challenge for generating design probes because it is technically three models: a pretrained general language model, its distillation, and the fine-tuned version optimized for a sentiment analysis task. We found that this challenged the SMC paradigm, since it was not clear how much information to include (if any) of the prior base models. To explore this problem, we decided to use the text from \textit{both} the DistilBERT and SST-2 fine-tuned model cards\footnote{\href{https://huggingface.co/distilbert-base-uncased-finetuned-sst-2-english}{distilbert-base-uncased-finetuned-sst-2-english} commit \texttt{03b4d19}}. 



\vspace{1mm}
\noindent\textbf{Design Probes and Interactive Model Card Concept.} We developed design probes~\cite{Wallace:design_probes:2013} of an SMC, RGR, and IMC. For the SMC, we used the all the text from the reference model card, with some exceptions, laid out according to the specifications in Mitchell~\textit{et al.}~\cite{mitchell10.1145/3287560.3287596}. Because the reference model card does have some interactivity (the ability to test your own sentence), we included that as well. An RGR unifies four common evaluation paradigms including performance on data subpopulations, which we explore in this concept study.  We developed an RGR probe using the built-in subpopulations to which we added a custom set of potentially sensitive subpopulations (i.e., race, gender, etc.).

Our IMC concept~(\autoref{fig:phaseI_concepts}) includes redeveloped elements of the SMC and RGR.
We proposed adding interactive elements (shown with green border) that would allow users to upload their own data (\texttt{data upload panel}), add their own sentences (\texttt{example generation panel}), and/or define new subpopulations from the model's training and test datasets for exploration (\texttt{example generation panel}). We also include an interactive data comparison visualization (\texttt{data comparison panel}) that shows the sentence embeddings of the model's data together with the analyst's data. We implemented the design probes as paper prototypes using Google Slides. For the IMC we mocked up interactions through slide links, transitions, and animations. Text for these paper prototypes was taken verbatim, with some small exceptions, from the existing \texttt{DistilBERT} and the \texttt{SST-2 fine-tuned} model cards.



\vspace{-3mm}
\subsection{Study Procedures}\label{study_one:general_procedures}

\subsubsection{Study Design} 
Our study comprised semi-structured interviews using a grounded approach~\cite{muller2014curiosity} and incorporated conceptual design probes (Section~\ref{sec:studyone_probes}).
Study sessions were scheduled for 60 minutes, and participants were offered an honorarium of \$150 upon completion.
Full study protocols and materials are available in the online materials, 
while here we provide only a brief overview. During the study, participants were asked about their background and familiarity with model cards, Robustness Gym reports, or other types of model documentation. They were then shown examples of a standard model card, a Robustness Gym report, and our concept for an interactive model card (Section~\ref{sec:studyone_probes}). They were asked to explore each of the forms in a think-aloud protocol, and the study moderator also prompted them 
with additional questions related to their interpretations and needs for understanding the model.
As they considered each of these forms of model documentation, we asked participants to reflect on the strengths and limitations through two lenses: their own needs as experts and those of a non-expert analyst using a sentiment analysis model in their work (see the \href{https://osf.io/9d83t/?view_only=7359695185b04468a7ae4a2e189691c0}{online materials}).


\begin{wraptable}{r}{0.46\textwidth}
\vspace{-5mm}
\caption{Overview of roles and expertise of participants.}
\Description{A table that shows an overview of the ten expert participants in our first study. The table has five columns : a unique ID, the participant role (or job title), their area of expertise, their familiarity with standard model cards, and their familiarity with Robustness Gym reports. The summary statistics of this table are summarized in the main text.}
\vspace{-3mm}
\scriptsize
\centering
\begin{tabular}{p{1.5em}p{10em}p{4em}p{3.5em}p{3.5em}}
\toprule
ID & Role & Expertise & SMC\newline Familiarity & RGR\newline Familiarity \\ \midrule
E01 & User Research Director & Ethics & No& No\\
E02 & AI Ethics Lead & Ethics &Y es & Yes \\ 
E03 & Ethics Data Scientist & Ethics &Yes &No \\ 
E04 & Doctoral Student - CS & Auditability&Yes &No \\ 
E05 & Doctoral Student - ML & Developer &No& No \\ 
E06 & User Research Lead & Ethics &Yes &No \\ 
E07 & Senior Researcher - NLP & NLP & Yes& Yes\\ 
E08 & Postdoctoral Fellow - NLP/HCI & NLP &No &No \\ 
E09 & Policy Research Fellow & Ethics &Yes &No \\ 
E010 & Data Ethnographer & Ethics & No& No\\ 
\bottomrule
\end{tabular}%
\vspace{-3mm}
\end{wraptable}

\vspace{-2mm}
\subsubsection{Recruitment}
We set a target of 10 participants aiming to recruit at least one participant from each of the following categories: ML/AI development and design,  Ethics, and NLP.  We reached out to 23 participants, recruiting until we had met our target study size. Of the 13 participants who were not included in our study, five declined and the rest did not respond.  We recruited participants based upon their publication or professional history using a combination of personal connections and cold emails. All participants had exposure to documentation for ML/AI models in some form, but only six were familiar with model cards and two with RGRs.

\vspace{-2mm}
\subsubsection{Data Collection and Analysis}\label{study_one:qual_coding}
Data collection included observer notes and video recordings for participants who consented to recording. We recorded approximately 11 hours and 15 minutes of video, including the interview and an on-camera debrief between the study moderator and note taker. Video transcripts were automatically generated, including speaker identification, and were verified for accuracy by the authors. We conducted iterative \textit{thematic analysis}~\cite{charmaz2014constructing}, surfacing initial codes through debriefs and review of transcripts. The first author then conducted focused coding for all transcripts, and both first authors synthesized themes and categories. All authors also surfaced implications that informed the design of our \textit{IMC functional prototype} (Section~\ref{study_two:functional-prototype}).


\subsection{Results}
In~\autoref{tab:studyone_tab} we summarize five interconnected conceptual dimensions of participant perspectives on our probes. 

{\renewcommand{\arraystretch}{1.30}%
\begin{table*}[t!]
\caption{Themes from our concept studies with experts in ML/AI, NLP, ethics, and auditability. Support quotes for each theme are also presented in the interview context from which they were elicited: SMC = Standard Model Card; IMC = Interactive Model Card; RGR = Robustness Gym Report; All = general comments not directed toward any specific model card type.} 
\Description{A table with quotations that from the first study with ML/AI experts. The table has three columns, the last column has two nested columns. The first three columns are overall themes of the quotes, the second column has sub themes that emerged, and the last column has example quotations from the interviews. This last column is further broken down into two nested columns. The first is the context, indicates whether the participant's comment is toward the standard model card (SMC), robustness gym report (RGR), interactive model card (IMC), or all three (all). The second nested column contains the quote with the participants ID at the end. }
\vspace{-1mm}
\label{tab:studyone_tab}
\scriptsize
\centering
\begin{tabular}{p{8em}p{12em}p{3.5em}p{41.5em}}
\toprule
\textbf{Theme}     & \textbf{Subthemes} & \multicolumn{2}{l}{\textbf{Example from interviews}} \\
& & Context & Quote \\ \midrule
\multirow[t]{4}{*}{Stakeholders} & \multirow[t]{4}{11.5em}{Stakeholders, Language} 
&SMC&\textit{``Model cards look a lot more like a summary of an academic paper compared to getting  a more general understanding of model.''} [E07]  \\ 
& &SMC&\textit{``Obviously, this, this standard model card is fantastic for an engineer who's coming in and trying to build that model [...] but [I am] not sure what their client needs.''} [E09]  \\ 
& &All& \textit{``There’s a lot of jargon, some of it that can be inferred, but it’s not something that is immediately apparent or made explicit'' }[E10]\\ 
& &RGR&\textit{``[Precision, recall, accuracy], those are terms that are also going to be very confusing for someone with layman's background [...] I would also say that those terms are terms of art in many ways.''} [E09]\\  \hline

\multirow[t]{4}{7.5em}{Design Considerations} & \multirow[t]{4}{11.5em}{Information Design, Visual and Interaction Design, Implementation, Integrated Tooling}
%
%
%
&IMC&\textit{[Hovering over points in the data comparison chart] I would want to see sentences because it breaks the abstract representation into some concrete things''} [E08]\\
&&IMC&\textit{``It is hard to discover the interaction, so make it clearer what you get with the interactivity and make it clearer that interaction is possible.''}[E09]\\
%
& &SMC& \textit{``The developer has to make these model cards, but they are difficult to make, which is why these are not very informative.'' } [E08]\\ \hline

\multirow[t]{3}{7.5em}{Guidance} & \multirow[t]{3}{11.5em}{Actionability, Defaults, Education, Explanations } 
&All&\textit{``People who aren't trained in ML don't necessarily know what they're looking out for or what questions to ask of both the data and the model  [...] nudges could be useful.''} [E04] \\
& &All&\textit{``Supporting analysis that says how to interpret this would be much more useful, but just even if it's formatted nicely with definitions, it's still not going to be super useful. Help me understand why this matters.''}[E02]\\
%
%
& &All&\textit{``I wish that this [model card] would pull [me in more]. Tell me what to look at [...] maybe you could suggest the subpopulations that I might want to look at.''} [E01] \\
&&IMC&\textit{``How much education about concepts do you need to bake in here? Do you have an intro in the beginning of this interactive model card?''}[E02]\\\hline

\multirow[t]{3}{7.5em}{Trust and Safety} & \multirow[t]{3}{11.5em}{Bias, Ethics, Misrepresentation} 
&All&\textit{``If I really am sort of coming in new to [model cards], I want to understand what my responsibility is [...] when I'm engaging with this model. I want to understand what agency I have.''} [E01]  \\
%
%
& &SMC&\textit{``Does this model card seem like it's showing a well-rounded representative model? Or is it one that has some big, ethical or transparent issues?''}[E03]\\ 
& &All&\textit{``Who gets to decide how to build those systems? And what is the end use case for evaluation? So in the healthcare space, specifically, a lot of times people get caught up with the area under the curve score F1 metrics for evaluating models about how they impact people's lives.''}[E09] \\ \hline

\multirow[t]{4}{7.5em}{Sensemaking and Skepticism} & \multirow[t]{4}{11.5em}{Contextualization, Data Analysis, Examples, Information Seeking, Interpretation}
%
&RGR&\textit{``[I] might not pay as much attention to it [size of subpopulation]. And that really matters because you can see some of these examples--they have high scores, but then the [...] sample size is very small.''}[E04]\\
& &RGR, IMC &\textit{``Just give [people] some example sentences and let them decide how they want to search or what they want to search for. And that way they can see model weaknesses and capabilities.''}[E05]\\ 
&&SMC&\textit{``Some people read all the documentation straight; others, only when [...] they run into a challenge. And so there's a lot of information [in the standard model card] so, one question is [...] how many people are going to read through all this in advance when they're actually on the job and in practice?''}[E04]\\
\bottomrule
\end{tabular}
\vspace{-6mm}
\end{table*}
}


\subsubsection{Stakeholders}\label{study_one:res_stakeholders}
Without prompting, participants described additional stakeholders or students and the scenarios in which they share model cards or other documentation (e.g., related to model performance and tuning). One participant succinctly summarized that SMCs remained relatively untested with non-expert analysts: 
\vspace{-1mm}
\begin{quote}
\small
    \textit{``At least for the next couple of years, people are going to be using model cards for the first time. And these are going to be concepts that may be new for many individuals. And so giving them some hand-holding, as to `What does this mean, what am I supposed to do with this?' I think will be extremely helpful.''} [E02]
\end{quote}
\vspace{-1mm}

Several participants (n=6) articulated that the language and layout of the SMC appeared better suited for software or machine learning engineers. One participant suggested that the SMC seemed \textit{``very on the back end...like it feels like GitHub'' }[E01], which they felt would not resonate with stakeholders who are using, but not implementing, models. Moreover, three participants reflected that some of these specialized terms of art (i.e., accuracy, precision, recall) can be difficult for even students of ML and NLP; one participant said their students needed a `cheat sheet' of the terminology.

All participants expressed doubt that a non-expert analyst would understand the model card or know how to act on the information it contained.
One participant suggested  that an IMC could be more accessible to individuals across different roles, particularly if the design supported different information architectures or prioritization of information.  In contrast, another participant [E07] was emphatic that standardization was a critical component of model cards and should be preserved. They argued that interactivity adds subjectivity to model cards that should be visually distinct.

\vspace{-1mm}
\subsubsection{Design Considerations}
Participants critiqued the information design of all model cards; for the RGR and IMC, they also offered feedback on visual and interaction design. Most (n=6) participants identified overly technical language and jargon as a key failing of all of the model cards, and they proposed alternative language to capture the same meaning in more accessible ways. All participants said the amount of text in the standard model card made it difficult to quickly skim and understand the model. 
Most (n=8) participants expressed that the visual design and layout of the IMC helped their  \textit{``eyes [...] focus on information that [they] missed before''} [E04] compared to the SMC and RGR. However, a sense of focus/overwhelm varied among participants, and one participant articulated that the IMC contained \textit{``too much information and is very cluttered''} [E04]. Four participants stated that a summary of the model's purpose (i.e., ``this is a sentiment analysis model'') was needed and that it would improve the readability of the model card. Relatedly,  participants also pointed out areas in the RGR and IMC where they thought different information should be prioritized or emphasized (e.g., changing the color of the size column on the rightmost side of the RGR or emphasizing it in another way). Lastly, many participants suggested adding or emphasizing higher-level metrics about performance (e.g., making overall performance visually distinct from disaggregated performance).

\vspace{-1mm}
\subsubsection{Guidance} 
Participants indicated that various stakeholders would need help making sense of the information in the model card. Six participants emphasized the importance of providing specific guidance to help participants understand the information on the model card. 
Some forms of guidance suggested included
providing definitions, or visual cues in the form of nudges and warnings. One participant articulated that explainers for visualizations would be helpful because \textit{``it's pretty but..but what would I do [with it?]''} [E03]. One participant saw the direct potential for using interaction to provide guidance, suggesting that it \textit{``would help me [...] dig deeper''} [E01] into the model details.

Other forms of guidance were complex and geared more toward \textit{``help[ing] me understand why this matters [and] how does it help me make more informed decisions?''} [E02]. Supporting interpretability is complex but nearly all (n=7) participants indicated that examples could provide some useful guidance. However, having an effective default example to prompt end-users was also important 
[E04]. Another participant suggested that leveraging examples with interactivity could help 
non-expert analysts
\textit{``build a better mental model of how the model works and what it lacks''} [E05].

\vspace{-1mm}
\subsubsection{Trust and Safety}\label{study_one:trust_safety}
Ethics and safety considerations 
were a top concern.
The reference model card (Section~\ref{sec:studyone_probes}) had a short ethical considerations text, but many participants took exception with the training dataset being described as ``fairly neutral'' (\textit{``what does that mean even?''} [E08]).  One commented that Silicon Valley tends to apply a standard set of \textit{``moral values application across global and cultural contexts''} [E07], which diminishes the value of such ethical statements and assessments. Increasingly, such statements are seen as a \textit{``cop-out because it just kind of says `there is some bias somewhere' which is not entirely helpful.''} [E06]. Two participants suggested stating explicitly whether bias assessments had been conducted, and one suggested raising a warning if such bias analyses are not run. Several (n=4) participants expressed that it could be beneficial to add sensitive populations to the disaggregated model performance, but they suggested caution since the definition of these groups was culturally and geographically contextual. Moreover, they 
emphasized the need to prominently display
the total size of these subpopulations (and others) to support valid interpretation. 
For example, one participant initially thought the model performed well for a particular subpopulation, then saw the small sample size and revised their interpretation. 
Participants indicated that for these issues and overall, greater attention to information, visual, and interaction design would improve trust and safety. 

\vspace{-3mm}
\subsubsection{Sensemaking and Skepticism} 
Some of the themes above converged 
as participants discussed the ways they believed non-expert analysts 
would explore and make sense of information in the model cards. 
Participants tied interaction positively to these themes, indicating that it would be beneficial to \textit{``allow domain experts to be able to interrogate the models and formulate rules that are semantically meaningful to test against the model'' [E08]}. Importantly, the interactivity would enable people to \textit{``elicit more questions'' } [E04] that allowed them to form and test their mental models about the data and better \textit{``see the things that model lacks''} [E05].

\vspace{-3mm}
\subsection{Summarizing Design Guidance for an Interactive Model Card}\label{study_one:design_guide}
From our findings in this first study, we distilled a set of design guidelines (\textbf{DG}) to inform our functional IMC prototype.
Based on what we learned from experts, these guidelines have broad implications for the documentation of AI/ML models and would be applicable beyond the task presented in the design probe.

\noindent\textbf{DG1: Give careful deliberation to the design of information hierarchies, representations, and interactions.}
Prior work related to SMCs has focused primarily on the categories and depth of information that model documentation should include. However, to be sure that this content is effectively understood by a wide audience, the hierarchy and design of this content is equally important. Without this consideration, a model card may fail in its goals. Our conceptual study emphasized the importance of information design within both standard and interactive model cards. This includes the judicious use of language, visual cues, layout, data visualization, and interaction. Critically, our study cautions that the act of information design must be deliberate and, we argue, testable to be sure it is accurately interpreted.




\noindent\textbf{DG2: Use interaction to help users develop conceptual understandings.}
The reference model card primarily used interaction as a preview of the model's outputs. However, experts saw the opportunity to use interaction as a way to help non-expert analysts develop their conceptual understanding of the NLP model. 
Interaction modalities must balance the subjectivity of contextual explorations with objective details about the model and its performance, so they should be designed and evaluated accordingly.
The final consideration for interaction is that it must be  discoverable and supported with defaults that help people orient to the affordances. 

\noindent\textbf{DG3: Scaffold important information with actionable guidance.} Additional guidance will be necessary to support interpretability of the outputs. SMCs have been primarily tailored toward developers and machine learning experts, 
which can cause challenges for non-expert analysts who need to interpret the documentation for their own contexts.
Clear guidance can support non-experts in understanding and knowing how to act on the information.
Warnings, prompts, and summaries are all examples of guidance that participants suggested could help guide analysts in making sense of 
model cards.
Visual cues can also
support the overall architecture and flow of information in the model card. 

\noindent\textbf{DG4: Implement defaults that promote productive skepticism.} Not all readers of a model card will want to interact with it. In addition to making interaction discoverable, it should also be enticing and promote a productive skepticism about what the model does and how it can be used. We argue that this productive skepticism would make ideas of intended and unintended uses more intuitive than simply being told what they are. Choosing appropriate default information is important to encourage this kind of engagement with a model card. We propose sensitive populations as a starting point and we used the Robustness Gym technology~\cite{goel2021robustness} to automatically implement these defaults.


\vspace{-2mm}
\section{Study 2 : Evaluation with non-expert Analysts}\label{sec:study_two}
We used the design guidelines from the initial concept study to implement a functional prototype for an interactive model card (IMC) that we assessed with 20 non-expert analysts. 

\vspace{-3mm}
\subsection{Interactive Model Card Functional Prototype}\label{study_two:functional-prototype}
\begin{figure}
    \centering
    \includegraphics[width=0.97\textwidth]{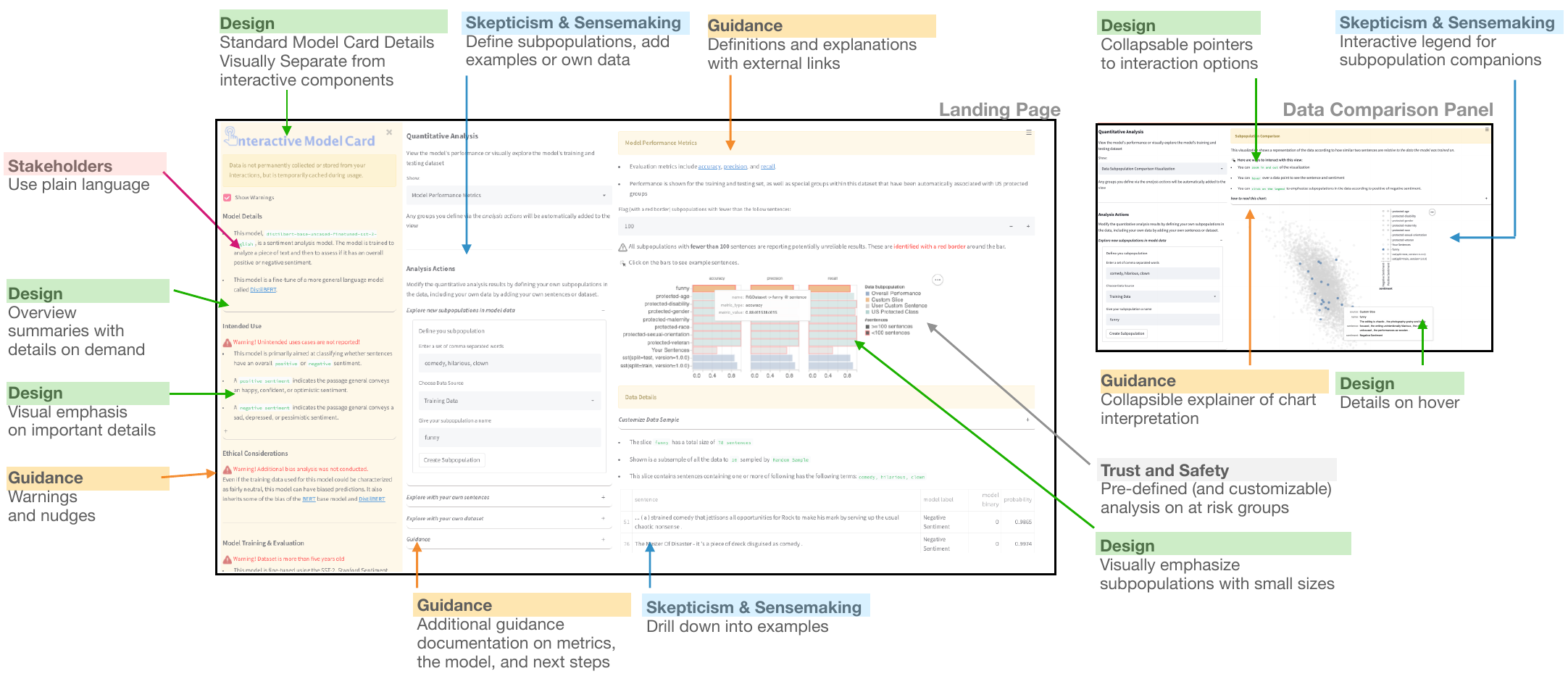}
    \vspace{-3mm}
    \caption{IMC Functional prototype that we presented to participants in the second study. Overlain are some examples of how feedback from the first study influenced the prototype design. The code for this prototype is available in the online materials.}
    \Description{An image of of the functional prototype of the interactive model card. Two panels are shown. The larger panel shows the main view of the interactive model card, while the second smaller panel shows the data comparison. The emergent themes that influenced the model card design are overlain. Each theme is shown as a color coded rectangular box with an arrow indicating the specific element of the functional prototype that was influenced by that theme.}
    \label{fig:phaseII_imc_concept}
    \vspace{-5mm}
\end{figure}

The functional prototype of the IMC 
reifies the design guidelines; in ~\autoref{fig:phaseII_imc_concept} we show 
our IMC with themes from~\autoref{tab:studyone_tab} overlain to emphasize the sources of design guidance. 
We visually separated information between a model overview component--reference model card information--and the `contextual' component, where interactivity is used to probe the model's performance and underlying data.  Within these components, we introduced elements of information and visual design to present the content (\textbf{DG1}) and create avenues of interactive engagement (\textbf{DG2}). 
Across both components, we use font faces, color, and highlighting to emphasize important content like the model's task (i.e., sentiment analysis) and its range of outputs. 
We also introduce two levels of information hierarchy. The first level summarizes vital information as a bulleted list; the second level
provides
technical details that are only visible when the analyst expands the content. In the overview component we also reduced the jargon from the SMC text. 

We use interaction to support concept building, sensemaking, and skepticism around the model and its performance (\textbf{DG2}). 
The interactive data visualizations 
show the model's performance and explore the model's data. The performance visualization shows the overall train and test data set performance according to accuracy, precision, and recall (the reference model card only shows accuracy). Using Robustness Gym~\cite{goel2021robustness} we developed a set of terms that 
relate to
subpopulations of US protected classes (e.g., race, gender, veteran status; \textbf{DG4}) that are also visualized
We introduce three ways for analysts to further probe model performance: defining custom subpopulations within the model's data, adding their own sentences, and adding their own dataset. When an analyst chooses to add a sentence, they will see a summary of the model's sentiment prediction and have the option to contest the result; the `sentiment label' for the new sentence is based on the participant's choice.  We have also defined a set of default sentence templates of mixed sentiment~\cite{potts2020dynasent} and with sensitive attributes~\cite{czarnowska2021quantifying} to help analysts further explore examples that are distinct from the model's data. Finally, the data visualizations are updated in real-time to reflect the model's performance in these newly defined subpopulations or data (sentence or dataset).

We scaffold the information in the model card with guidance that supports the interpretation of 
textual
information and visualizations~(\textbf{DG3}).  Examples of guidance include simple statements explaining how to interpret visualizations or interact with them, definitions of terms, and 
instructions for adding new data.
Lastly, we include a `Guidance' section that includes more general information about sentiment analysis models and the interpretation of performance metrics. We 
incorporate
warnings to alert the analyst to potentially unreliable or incomplete information in the model card. The performance visualization also alerts analysts 
when a subpopulation's size is lower than a specified threshold.

The IMC is implemented in Python 
using \texttt{streamlit}\footnote{\url{www.streamlit.io}}, with visualizations and data interactions supported by \texttt{Altair}~\cite{vanderplas:altair:2018}. \texttt{Robustness Gym}~\cite{goel2021robustness} and \texttt{Gensim}~\cite{rehurek:gensim:2011} are primarily used to handle data inputs and outputs between the cards visual components, and are supported by other packages (\texttt{nltk}~\cite{bird:nltk:2009}, \texttt{numpy}~\cite{harris:numpy:2020}, \texttt{pandas}~\cite{mckinney:pandas:2010}, \texttt{sklearn}~\cite{scikit-learn}).


\subsection{Study Procedures}\label{sec:study2_procedures}


\subsubsection{Study Design} We conducted a 60 minute semi-structured interview with each of the participants,
asking them to describe
their use of ML or NLP models as well as the type of model documentation they consume or, if applicable, produce. Participants were then introduced to the format of a model card and directed to the DistilBERT SST-2 fine-tuned reference model card (Section~\ref{sec:studyone_probes}). They were asked to explore this model card and think out loud about what information they would want to gather, whether it was available, and 
how they would wish to see this information. Next, participants were given a demonstration of the interactive model card and were asked to explore it as they had the reference model card. In the final 15 minutes of the session, participants were asked to consider the IMC with respect to its usability, functionality, safety, and ethics. These question prompts were developed based on the findings of the previous study and a set of explainability characteristics
that we adapted from those
described by Sokol and Flach~\cite{sokol10.1145/3351095.3372870}.
We also asked participants to compare the IMC to the reference model card. At the conclusion of the study participants were compensated \$125.

\vspace{-1mm}
\subsubsection{Recruitment} We set a target of 20 participants and recruited using the \href{https://userinterviews.com}{User Interviews} platform. We screened eligible participants using a questionnaire and  \textit{a priori} inclusion and exclusion criteria. In total 176 participants responded. A total 117 participants were excluded, in some cases because they indicated never using ML or NLP models (n=89), while others (n=28) had graduate or undergraduate training in ML, statistics, or computer science. The remaining 59 candidates were separated into two groups according to the frequency with which they use code-based tools (i.e., Python, R, etc). Type I (n=35) analysts do not use programming tools at all or very infrequently (less than once a quarter), whereas Type II  (n=24) analysts report daily or weekly use.  
We recruited an even number of participants from both categories of analysts (\autoref{tab:study_two_participants}), anticipating each group might offer a different perspective to our study.Participants were compensated \$125

{\renewcommand{\arraystretch}{1.15}%
\begin{table}[t!]
\caption{Overview of Participants for the Second Study.}
\Description{A table summarizing the demographics of participants in our second study. The table was 8 columns. The first two columns concern the study details, with one column providing the participant's study ID and the second showing the category of analyst they are (Analyst I or Analyst II). The next four columns give background details on the participant; the columns in order have the participants' regions (Canada, US, UK, or others), their role (or job title), their industry, and the size of their organization. The final two columns concern the participants skills and proficiency with ML/DS and sentiment analysis, respectively. }
\vspace{-3mm}
\scriptsize
\begin{tabular}{p{2.5em}p{4em}p{7em}p{13em}p{12em}p{5em}p{5em}p{4em}}
\toprule
\multicolumn{2}{l}{\textbf{Study Data}} & \multicolumn{4}{l}{\textbf{Background Details}} & \multicolumn{2}{l}{\textbf{Skills and Proficiency}} \\ 
ID & Category & Region & Role & Industry & Org. Size & ML/DS\newline Proficiency & Sentiment \newline Analysis\\ \midrule
A1-01 & Analyst I& United States &Business and Marketing Analyst& Media &10000+& Basic& No\\
A1-02 & Analyst I& Canada & Project Manager & Finance &5001-10000& Basic& No\\
A1-03 & Analyst I& United Kingdom &Product Analyst& Insurance &201-1000& Limited& Yes\\
A1-04 & Analyst I& United States &Finance Director& Finance &10000+& Limited& Yes\\
A1-05 & Analyst I& United States &Product Analyst& Finance &10000+& Basic& No\\
A1-06 & Analyst I& United States &Business Analyst & Information Technology &1001-5000&Advanced& No\\
A1-07 & Analyst I& United States &Project Analyst& Environmental Services &1001-5000& Basic& No\\
A1-08 & Analyst I& United Kingdom &Automation Developer& Information Technology &201-1000& Basic& No\\
A1-09 & Analyst I& Canada &Business Data Analyst& Finance &5-200& Basic& Yes\\
A1-10 & Analyst I& South Africa &Business Intelligence Analyst& Finance &1001-5000& Basic& No\\  \hline
A2-01 & Analyst II& United States &Data Scientist & Government&10000+& Advanced& Yes\\
A2-02 & Analyst II& United Kingdom &Analytics Manager & Marketing \& Advertising &51 - 200& Advanced & Yes\\
A2-03 & Analyst II& United States &Machine Learning Researcher& Material Engineering &1001-5000&Advanced & Yes\\
A2-04 & Analyst II& United States &Machine Learning Developer& Information Technology &51-200& Advanced& No\\
A2-05 & Analyst II& Canada &Business Data Scientist& Telecommunications &10000+& Advanced& Yes\\
A2-06 & Analyst II& United Kingdom &Business Analyst& Finance &1001-5000& Advanced& Yes\\
A2-07 & Analyst II& United Kingdom &Data Analyst& Information Technology &201-1000& Advanced& Yes\\
A2-08 & Analyst II& United States &Data Engineer & Education &51-200& Advanced& Yes\\
A2-09 & Analyst II& United States &VP Data \& Analytics & Education &51-2000& Advanced& No\\
A2-10 & Analyst II& Canada &Analytics Solutions Developer& Retail&1001-5000& Expert&No\\
\bottomrule
\end{tabular}
\vspace{-5mm}
\label{tab:study_two_participants}
\end{table}
}


\vspace{-1mm}
\subsubsection{Data Collection and Analysis}
Data was collected and analyzed in the same manner as reported in Section~\ref{study_one:general_procedures}. In this study we recorded approximately 23 hours and 30 minutes of video, which included both the interview and an additional on-camera debrief between the study moderator and notetaker. We used the themes surfaced in the first study to seed this analysis.


\subsection{Results}
The reference model card (Section~\ref{sec:studyone_probes}), here acting as the SMC, and the IMC each had features that were 
preferred 
by both types of analysts. However, overall the IMC was favored over the SMC, with participants echoing the concerns that were previously articulated by the experts in Section~\ref{sec:study_one}. A summary of participant responses is shown in~\autoref{tab:studytwo_eval}. Compared to the prior study, non-expert analysts provided greater depth towards the themes of \textit{stakeholders}; \textit{design considerations}; and \textit{sensemaking \& skepticism} in particular. This depth allowed us to expand on the subthemes from~\autoref{tab:studyone_tab} and 
broaden our conceptions
of how model cards 
might
be used. 
In comparison to experts, non-expert analysts contrasted the \textit{understandability \& interpretability} of information between the SMC and IMC with more concrete details, leading to compelling insights around design implications.
Additionally, participants' responses to our explicit prompts around usability, functionality, safety, and ethics (summarized in ~\autoref{tab:studytwo_prompts}) allowed us to expand upon dimensions surfaced in our first study. They also helped us contextualize participants' overall impressions of the IMC, summarized in~\autoref{tab:studytwo_eval}.
We did not observe large differences in patterns of responses between the two analyst groups.
In Section~\ref{sec:study2-findings-overall}, we outline participants' overall impressions of the IMC. After that we focus on three themes that 
highlight
the importance of human-centered design in making model cards useful and usable in organizational settings.

\subsubsection{Evaluating our Design Choices}\label{sec:study2-findings-overall} 
{\renewcommand{\arraystretch}{1.30}%
\begin{table*}[t!]
\caption{Themes from our evaluative studies with non-expert analyst. The reference model card represents the SMC in this study.}
\Description{A table with quotations that from the second study with non-ML/AI experts experts. The table has three columns, the last column has two nested columns. The first three columns are overall themes of the quotes, the second column has sub themes that emerged, and the last column has example quotations from the interviews. This last column is further broken down into two nested columns. The first is the context, indicates whether the participant's comment is toward the standard model card (SMC), robustness gym report (RGR), interactive model card (IMC), or all three (all). The second nested column contains the quote with the participants ID at the end. }
\vspace{-3mm}
\label{tab:studytwo_eval}
\scriptsize
\centering
\begin{tabular}{p{7.5em}p{12em}p{3em}p{43em}}
\toprule
\textbf{Theme}     & \textbf{Subthemes} & \multicolumn{2}{l}{\textbf{Example from interviews}} \\
& & Context & Quote \\ \midrule
\multirow[t]{4}{7.5em}{Stakeholders} & \multirow[t]{4}{12em}{Community, Language, Sharing, Teams} & SMC, IMC & \textit{``Overall I am really impressed with the language clarity that is there right now, compared to [the reference model card]; the biggest difference is its clarity.''} [A2-09]\\
& & IMC &\textit{``We actually have had a lot of communication issues in the team  between our data science guru and the rest of the business and I think this actually might be really helpful''}[A2-02] \\
 & & IMC & \textit{``If you're trying to bring in a machine learning tool or a machine learning process, you've got to go through so many hoops to get that sign off, this is a sort of thing that would just be a godsend to have''} [A1-08]\\
 & & IMC & \textit{``This [IMC] is something that I could send in a pinch to any stakeholder. So if I was in a meeting with a whole bunch of stakeholders [...] and somebody's like, `Hey, what is the model? What does it do?' You know, I could answer most of these questions. But in a pinch, I'd say like, here's a link, go interact with it.''} [A2-04]\\ \hline
\multirow[t]{4}{7.5em}{Design} & \multirow[t]{4}{12em}{Assessment \& Validation, Comparison, Defaults, Information Priority,  Interaction, UX Patterns, Visualization} & SMC & \textit{``The first thing I am looking for is an introduction to the model and what it does and what it is used for. I don’t know if I see it here''} [A2-10]\\
& & IMC, SMC & \textit{``So, interactivity is the main one. This is more of an all or one, whereas the [SMC] makes me use another tool. It [SMC] is more like a shopping mall, where I need build it [the analysis] on my own at home.''} [A1-03]\\
& & IMC, SMC & \textit{``You want it all accessible, the fact that it’s one page is great. Like with the other one I had to click around a lot to get the information, but this model card doesn’t have that problem, which is great.''} [A2-02]\\ 
& & IMC & \textit{``It gives me a good outline of what I am looking at. The few examples, short, sweet, and to the point are helpful.''}[A1-07]\\
&& IMC & \textit{``Maybe it’s my time working in finance, but bullet points work for me''} [A2-02] \\ \hline
\multirow[t]{4}{7.5em}{Sensemaking \newline\& Skepticism} & \multirow[t]{4}{12em}{Awareness, Confidence, Contestability, Examples, Exploration, Prompts, Warnings, Risk Group Analysis} &  SMC, IMC  & \textit{``It [SMC] seemed more oriented towards somebody who wants to, like develop and train it [the model] and deploy it [...] this [IMC] does give me the same information, like it gives me the same context [...] but it's also giving me the tools to kind of look at the model on a higher level.''} [A2-04] \\ 
& & IMC &  \textit{``I absolutely like this quantitative analysis [...] which lets me play around, and I can see whether the model is working or not. So I'm not just taking the the word of the developer of the library''} [A2-09] \\
& & SMC, IMC & \textit{``if I ever see something that just reports one metric [...] I'm always like a little bit skeptical that maybe they just picked that up because that's what looks good''}[A2-01]\\
& & IMC & \textit{``I do like the fact that you can [...] agree or disagree [with the model's prediction], because sometimes it's nice to have that ability} [A1-06] \\ \hline
\multirow[t]{4}{7.5em}{Understandability \newline \& Interpretability} & \multirow[t]{4}{12em}{Accessibility, Clarity, Data Understandability, Guidance, Model \& Result Interpretation, Information Overload, Information Sufficiency, Trust} & IMC & \textit{``This gives me more assurance that this is the right model or not''} [A2-09] \\
& & SMC &  \textit{``I can tell it, it's used for some text classification[...] But, in terms of what this thing does, it's not obviously intuitive enough, unless you've caught some context around it. So I filled in a lot of the blanks.'' }[A2-08] \\
& & SMC & \textit{``What is the sentiment treebank [SST-2 dataset]? Like, is it a list of sentences? Where did they come from? Who generated them? Because that was also what I was confused''} [A2-01] \\
& & SMC & \textit{I feel that it’s too much information on this model. If, maybe, could I  have a better summary?} [A1-10]\\
& & SMC & `\textit{`The most important bit is what the purpose of the model is, and it’s good at the top, otherwise I didn’t know what what I was looking at, I had to guess how it actually works and what is trying to tell me''} [A2-05]\\
\bottomrule
\end{tabular}
\vspace{-5mm}
\end{table*}
}

Nearly all participants (n=18) 
found the IMC easier to read and understand (with some caveats discussed below). One participant said that the IMC showed information that \textit{``allowed [them] to understand what this model actually is supposed to be used for'}' and \textit{``this is a lot better at level setting''} [A2-08]. 
Participants articulated that several components of the IMC helped them engage with the model and data more effectively 
by clarifying and prioritizing information, 
particularly:
layout of information (n=13); language (n=6); ability to add examples (n=15); and visual and interactive elements (n=16).
Three participants voiced the importance of being able to contest the model's results, and wanted to see this ability expanded beyond adding sentences (Section~\ref{study_two:functional-prototype}) to the model's training and testing data or uploading their own dataset. One participant stated that the ability to \textit{``agree [or] disagree is kind of a check and balance''}[A1-06]. Nearly all (n=16) participants also made suggestions for improvements to the IMC design. These included changes to elements of the interface that were confusing (n=5), the layout (n=6), and types of guidance (n=1) provided. 
A few participants (n=4) expressed that 
text, visuals, and interaction options led to information overload.
Participants suggested that 
integrating the model's code and files, present in the SMC, would be valuable for the IMC.

When asked, nearly all participants (n=15) indicated that the IMC was something they would use in their daily work, while some (n=2) indicated it was context-dependent.
Participants
reported that the IMC, with or without interactivity, was preferable to the SMC because the information was easier to understand. Notably, the addition of guiding elements, suggested in the previous study~(\autoref{tab:studyone_tab}), were well received and did help participants understand and interpret the model card information. Many of these non-interactive elements, which include visual cues, warnings, and prompts, among others, can be used to augment SMCs without having to add interactivity. 
However, participants indicated that interactivity added substantially to the SCM
because it allowed participants to \textit{``play with the data and see the results [in a way that] that feels super intuitive instead of explaining everything''} [A1-09]. 

{\renewcommand{\arraystretch}{1.30}%
\begin{table*}[t!]
\caption{Examples of participant responses on the usability, functionality, safety, and ethics of the IMC.} 
\Description{A table with three columns that summarizes participants response to specific prompts on  usability, functionality, safety, and ethics of the IMC. The first column indicates the prompt, the second has the definition of the prompt, and the third has example quotes from the second study. }
\vspace{-3mm}
\label{tab:studytwo_prompts}
\scriptsize
\centering
\begin{tabular}{p{4.5em}p{10em}p{50em}}
\toprule
\textbf{Prompt}     & \textbf{Definition} &\textbf{Example from interviews} \\
\midrule
\multirow[t]{3}{4em}{Usability} & \multirow[t]{3}{10em}{The efficacy of the design choices to help assess and digest the information in the model card} &  \textit{``This feels like plain English [...] like a very simple way of saying, `hey, this is what we have. this is what it does.' '' }[A1-01] \\
&& \textit{``I think this [IMC] tells a better story, [...] and this tells me, what's the model? What's it supposed to be used like?''} [A2-08] \\
& & \textit{``It's not only the numbers, but it's also the insights and  visually being able to see[...] what those numbers are saying''} [A2-04] \\
& &  \textit{``The visualization, as far as the reporting goes, it has got a good description of what all that means. A lot of the help text and description text is there.'' }[A1-06] \\ \hline
\multirow[t]{3}{4em}{Functionality} & \multirow[t]{3}{10em}{The ability to use the IMC to assess the relevance and applicability of the model to their routine work}&  \textit{``This would be perfect for internal stakeholders, like technical stakeholders, direct managers, CTO, data science team, you know, people like that. I would want a simplified version of this for external stakeholders. But I also have all the data here and it looks like I can extract it if I want to.''} [A2-04] \\ 
&&\textit{``If I had a tool like this, I would start to explore, and pick out trends or phrases, in their data, so that I can start to see that there are opportunities from a features development perspective.''}[A1-03]\\
& & \textit{``I've got a data set that I want to run sentiment analysis on, I'm going to use this model, drop it in through my CSV, boom, now I want to see the visualizations great. And they're going to look really good when I'm presenting to my boss.''}[A1-03]\\
&&\textit{``It [the model card] would be more about ruling out that it's not the right model, you know, and then I might have like six or seven model cards open, as I'm going through all the models, I can find, like, here's my list of things we should try.''}[A2-07]\\ \hline
\multirow[t]{4}{4em}{Safety} & \multirow[t]{4}{10em}{The ability to assess safety risks for a model, including risks for security, privacy, robustness, and related dimensions} &  \textit{``I think it [the IMC] gives you the basis to have that conversation.''}[A2-02] \\
& &\textit{``It is the person’s responsibility to make sure they know what they are doing and not someone who is providing the model to them.'' }[A2-06] \\ 
& & \textit{``The safety risk is more in the problem itself and not the model you are applying.''} [A2-07] \\ 
& & \textit{``With the protected classes I think it’s a good feature to have, I think it’s one of the more important ones. Protected classes make me think that things are headed in the right direction. Also I like the warnings.''} [A2-08] \\
\hline
\multirow[t]{4}{4em}{Ethics} & \multirow[t]{4}{10em}{The ability to assess ethical consequences encompassing things like potential harm to vulnerable people or ecosystems, unintended bias, or similar issues}  &  \textit{``I think it's useful as well to say, `You know, what, you're gonna have to go and do your own research' [...] you can't just say `Oh, just go and read this.'[Use the] model card to assist you, but you're going to have to go and do a bit of learning yourself.''} [A1-08]\\
&&\textit{``I don’t think it helps me with ethical consequences, but it helps to give me some considerations.''} [A1-03]\\
& &\textit{``I think also having these categories [protected classes] made me curious and think about it.''} [A1-09]\\
&& \textit{``Bias is also like a term of art in machine learning, right? So it's like, I read it as bias, like, Oh, it's just talking about bias from the standpoint of overfitting. It has nothing to do with like, discrimination and ruining people's lives.''}[A2-07] \\
\bottomrule
\end{tabular}
\vspace{-6mm}
\end{table*}
}


\subsubsection{Model Understandability - Getting the Basics Right}
One of the most salient 
implications
of the IMC design was that it helped participants accurately understand what the model was and what it did. In this study, we were surprised to discover that, when using the SMC, it was hard for participants to articulate \textit{that the model performed sentiment analysis}. While experts had expressed that non-expert analysts would have some difficulties interpreting that information, this finding went beyond our expectations. Overall, 
almost half of the
participants were not able to provide the correct interpretation of what the model did using the SMC alone. Among these participants, four thought it was a general text classification model, but could not anticipate its outputs (positive or negative sentiment classification), two thought that DistilBERT and its fine-tuned version were both sentiment analysis models, one participant did not answer (\textit{``I would have a tough time understanding what it does.''}[A2-08]), one thought that SST-2 dataset was the model but could not articulate what the model did. Even three participants who did correctly describe the model were not confident in their interpretation : \textit{``I would guess that it's sentiment analysis based on text, but I really don’t know''}[A1-03]. 

One source of confusion was the need to consume information across two model cards (DistilBERT base model and SST-2 fine-tuned).
Guided by our earlier study, we tackled this challenge directly and used the design elements of the IMC to draw attention to what the fine-tuned model did and how it differed from the base model. The response of participants validated this 
approach, with one stating that \textit{``the side panel was exactly what I was looking for and it highlighted some of the words, then I know that this is something that I might be interested in looking at'' }[A2-06].

\vspace{-2mm}
\subsubsection{The Role of Model Cards in Sociotechnical Systems} 
Nearly all (n=18) participants described, without being prompted, how the IMC would help them share information with internal and external stakeholders. The actionability of a model card was frequently (n=13) tied by participants to its ability to be shareable and understandable to others with whom they work.
Moreover, participants discussed 
interwoven social and technical
processes and described where either an IMC or SMC would fit within them. 
Participants described how the model details (overview component in the IMC, Section~\ref{study_two:functional-prototype}) could help them with an initial assessment of the model's validity for their use case. A quick answer is essential, and one participant stated,  
\textit{``I would lose patience after 30-40 seconds if I have to put a lot of effort into what I'm looking for''}[A2-06].
If the participants were convinced the model was suitable for their use case, they would move on to the interactive components to explore it further. One benefit of this interactivity was that it allowed participants to conduct lightweight experimentation, which was cumbersome using the SMC because \textit{``you [had to] deploy the model in Python yourself, put in your own sentences''} whereas with the IMC \textit{``just being able to do it like this...ease of use is off the charts''} [A2-02]. Finally, the model card would be used to start a dialogue with other stakeholders. These stakeholders included executive or technical team members, including data scientists.

Beyond integrating model cards with their social organization, analysts (n=7; the majority belonging to group II) saw the benefits of incorporating model cards into their existing technical infrastructure: \textit{``A huge win would be getting this to integrate seamlessly with open-source frameworks'' }[A2-01]. Here, the reference model card contained information and actions to obtain the model code and launch it --- something that was absent in the IMC. Moreover, three participants expressed that IMCs, if they could be easy to produce, would lower their existing model documentation burden.

\vspace{-2mm}
\subsubsection{Ethics and Safety are Challenging Topics} 
Participants had difficulty contextualizing the safety and ethical dimensions of either the SMC or IMC. 
Although the IMC included templates to examine ethically challenging examples (Section~\ref{study_two:functional-prototype}),
only four participants expressed that these features could stimulate discussion. The unintended uses of the model were similarly inaccessible, with one participant stating that \textit{``unintended uses are important [but] how would I know what [is] unintended?'' }[A1-01]. 
One participant expressed doubt that any model documentation could really give an adequate picture of ethical and safety risks, emphasizing that `\textit{`this [model card] would not be my only data point or else, you know, that's a very quick way to lose your job''} [A1-04]. Although the goal of model cards is not to be a substitute for further analysis of ethical and safety risks, this participant's comment echoes earlier observations from experts (Section~\ref{study_one:trust_safety}) that information about ethical risks is often dismissed. Others felt that topics of safety and ethics were too abstract, with one participant stating that they \textit{``wouldn’t worry too much about that to be honest, I guess it depends on the type of analysis you're doing.''} [A2-05]. 
While the IMC also provided template sentences to promote ethical thinking, participants largely favored their own examples.  Unexpectedly, two participants pushed back against the inclusion of more actionable ethical and safety content, articulating concerns that too much guidance could be detrimental and provide a false sense of security.

\vspace{-3mm}
\subsection{Summary}
We assessed our functional prototype for an interactive model card with twenty non-expert analysts drawn from across industry. Their feedback validates many of our design choices, but also offers avenues for further improvements. These non-expert analysts also illuminated a more detailed view of how model cards, more generally, might be used within the sociotechnical systems in their organizational contexts. Our findings point to 
the importance of information and visual design for transparency and interpretability of model cards.


\vspace{-3mm}
\section{Discussion}\label{sec:discussion}
Technology that enables low or no code data analysis is lowering barriers to developing and deploying deep learning models. As E02 astutely observed (Section~\ref{study_one:res_stakeholders}), the net effect is to broaden 
the field of
stakeholders who will need access to model cards in the future. 
Increasingly, the data workers taking on AI/ML development do not have formal education in the theoretical and technical underpinnings of these systems. We posited that, for this group, augmenting model cards with interactivity would make AI/ML systems 
more interpretable.

We
recruited and explored the perspectives of people who we believe are representative of the future data workforce. Our findings show that 
interaction in
model cards helped participants better understand 
model behavior and implications for their work.
Critically, the benefits of interaction were realized primarily through the careful architecture of information that included the choice of data visualizations, 
layout, language, visual cues, warnings and prompts. 
We contend that without these careful design choices, model cards are inaccessible to many individuals who use AI/ML models in their work. 
Our studies have also surfaced the value proposition data workers identify for model cards to support organizational decision-making and traceability.
From our collective findings we draw four key insights:

\vspace{1mm}
\noindent\textbf{1) Interactive model cards are a bridge to further analysis, not a substitute.} 
How interactive should a model card be? At what point does it cross over into an analysis interface?
In designing the IMC, we wrestled with these questions, and study participants were divided on the appropriate balance.
We argue that with thoughtfully designed interaction, model cards can be a first step for data workers to explore how trustworthy a model is and whether it might meet their needs, serving as a guide to deeper interrogation when desired.
For individual stakeholders, the most effective form for interactivity in model cards will be situated and contextual. 
However, future research could deepen our understanding of how particular design decisions impact the understandability and actionability of model cards.

\vspace{1mm}
\noindent\textbf{2) Interactivity and guidance bolster productive skepticism.}
Interaction supports individuals in contextualizing and interrogating model behavior (e.g., by adding their own data or contesting the model's results). The opportunity to \textit{``be really curious to just kind of play around''} [A2-05] can help mitigate anchoring bias~\cite{Nourani:anchoring:2021}. 
This sort of interactivity may encourage \textit{productive skepticism,} an orientation that is neither overly dismissive nor trusting.
Importantly, we found that scaffolding these interactions to support sensemaking requires clear and actionable guidance.
Our research has only scratched the surface of the dimensions and forms of guidance that help people understand and act on model behavior.
There remain many avenues to explore with IMCs as a foundation for this inquiry.

\vspace{1mm}
\noindent\textbf{3) Model cards should support the data work community, not just an individual.} Decision-making with data and deep learning models 
is a collaborative and distributed process involving information sharing over time and across organizational roles. 
For model cards to be adopted and impact organizational processes, they must support knowledge sharing and negotiation across  stakeholders with diverse backgrounds and perspectives.
Furthermore, the creation of interactive model cards need not be so burdensome. Our findings suggest that collaborative development, along with integrated programmatic tooling, could lower the burdens for model card generation, reuse, and refinement.

\vspace{1mm}
\noindent\textbf{4) Unintended uses, ethics, and safety are too ambiguous to be actionable.} Model cards are intended to surface, among other things, ethical and safety implications
-- a topic that is especially pressing for deep learning models. 
However, our research supports prior findings that these concepts are 
challenging to integrate into decision-making~\cite{sambasivan:datawork:2021,morley:what_to_how:2019}, 
and we identified some 
cynicism toward ethics statements, as well as 
assertions that ethical questions require contextual, situated examination.
Our findings indicate, unsurprisingly, that model documentation can spark ethical thinking, but will never substitute for it. 
We encourage further inquiry into other strategies toward fairness (e.g., metrics~\cite{czarnowska2021quantifying,thomas2020problem}), and we urge caution to ensure that these approaches are understandable and actionable for non-experts. 
Importantly, our analysis 
reinforces that considerations of ethics 
are personal and contextual, so they are unlikely to be addressed with a single approach.

\vspace{1mm}
\noindent\textbf{Limitations and Future Work}.
Design choices and the choice of sentiment analysis tasks
(Section~\ref{sec:studyone_probes}) 
likely influenced our findings.
We anticipate that specific IMC design elements (i.e., the choice of visual cues, data visualizations, etc.) will be refined and adapted in future work.
However, we believe that the 
broader design guidelines
(Section~\ref{study_one:design_guide}) and conceptual dimensions that we surfaced 
(Tables~\ref{tab:studyone_tab},~\ref{tab:studytwo_eval}, and \ref{tab:studytwo_prompts}) are robust and will generalize to other tasks and contexts. 
We also recognize that creating and learning how to use IMCs is a heavier burden than for text-based model cards. We have taken steps in this work to build toward adaptable and extensible model card generators. Future work will explore further strategies for reducing and re-distributing the work of model documentation.

%



    
\vspace{-3mm}
\section{Conclusion}
Together with ML/AI experts and non-experts, we co-developed a concept and functional prototype for an interactive model card for a large language model. 
Within the broader range of stakeholders using and deploying NLP and other ML/AI models, our work presents a timely and important examination of model cards from a human-centered design perspective. 
We foresee future opportunities for other researchers, practitioners, and developers to build from our findings to ensure that 
ML/AI models
are interrogable, contestable, and documented responsibly.

\begin{acks}
We wish to thank Britta Fiore-Gartland, Anna Bethke, Kathy Baxter, Wenhao Liu, Emily Witt, and Vidya Setlur for their thoughtful feedback on this work. We also gratefully acknowledge study participants for their time and insights.
\end{acks}

\bibliographystyle{ACM-Reference-Format}
\bibliography{main}


\begin{thebibliography}{77}


\ifx \showCODEN    \undefined \def \showCODEN     #1{\unskip}     \fi
\ifx \showDOI      \undefined \def \showDOI       #1{#1}\fi
\ifx \showISBNx    \undefined \def \showISBNx     #1{\unskip}     \fi
\ifx \showISBNxiii \undefined \def \showISBNxiii  #1{\unskip}     \fi
\ifx \showISSN     \undefined \def \showISSN      #1{\unskip}     \fi
\ifx \showLCCN     \undefined \def \showLCCN      #1{\unskip}     \fi
\ifx \shownote     \undefined \def \shownote      #1{#1}          \fi
\ifx \showarticletitle \undefined \def \showarticletitle #1{#1}   \fi
\ifx \showURL      \undefined \def \showURL       {\relax}        \fi
\providecommand\bibfield[2]{#2}
\providecommand\bibinfo[2]{#2}
\providecommand\natexlab[1]{#1}
\providecommand\showeprint[2][]{arXiv:#2}

\bibitem[\protect\citeauthoryear{Amershi, Cakmak, Knox, and Kulesza}{Amershi
  et~al\mbox{.}}{2014}]%
        {Amershi:Power:2014}
\bibfield{author}{\bibinfo{person}{Saleema Amershi}, \bibinfo{person}{Maya
  Cakmak}, \bibinfo{person}{W.~Bradley Knox}, {and} \bibinfo{person}{Todd
  Kulesza}.} \bibinfo{year}{2014}\natexlab{}.
\newblock \showarticletitle{Power to the People: The Role of Humans in
  Interactive Machine Learning}.
\newblock \bibinfo{journal}{\emph{AI Mag.}}  \bibinfo{volume}{35}
  (\bibinfo{year}{2014}), \bibinfo{pages}{105--120}.
\newblock


\bibitem[\protect\citeauthoryear{Amershi, Weld, Vorvoreanu, Fourney, Nushi,
  Collisson, Suh, Iqbal, Bennett, Inkpen, Teevan, Kikin-Gil, and
  Horvitz}{Amershi et~al\mbox{.}}{2019}]%
        {amershiGuidelines10.1145/3290605.3300233}
\bibfield{author}{\bibinfo{person}{Saleema Amershi}, \bibinfo{person}{Dan
  Weld}, \bibinfo{person}{Mihaela Vorvoreanu}, \bibinfo{person}{Adam Fourney},
  \bibinfo{person}{Besmira Nushi}, \bibinfo{person}{Penny Collisson},
  \bibinfo{person}{Jina Suh}, \bibinfo{person}{Shamsi Iqbal},
  \bibinfo{person}{Paul~N. Bennett}, \bibinfo{person}{Kori Inkpen},
  \bibinfo{person}{Jaime Teevan}, \bibinfo{person}{Ruth Kikin-Gil}, {and}
  \bibinfo{person}{Eric Horvitz}.} \bibinfo{year}{2019}\natexlab{}.
\newblock \bibinfo{booktitle}{\emph{Guidelines for Human-AI Interaction}}.
\newblock \bibinfo{publisher}{Association for Computing Machinery},
  \bibinfo{address}{New York, NY, USA}, \bibinfo{pages}{1–13}.
\newblock
\showISBNx{9781450359702}
\urldef\tempurl%
\url{https://doi.org/10.1145/3290605.3300233}
\showURL{%
\tempurl}


\bibitem[\protect\citeauthoryear{Arrieta, D{\'\i}az-Rodr{\'\i}guez, Del~Ser,
  Bennetot, Tabik, Barbado, Garc{\'\i}a, Gil-L{\'o}pez, Molina, Benjamins,
  et~al\mbox{.}}{Arrieta et~al\mbox{.}}{2020}]%
        {arrieta2020explainable}
\bibfield{author}{\bibinfo{person}{Alejandro~Barredo Arrieta},
  \bibinfo{person}{Natalia D{\'\i}az-Rodr{\'\i}guez}, \bibinfo{person}{Javier
  Del~Ser}, \bibinfo{person}{Adrien Bennetot}, \bibinfo{person}{Siham Tabik},
  \bibinfo{person}{Alberto Barbado}, \bibinfo{person}{Salvador Garc{\'\i}a},
  \bibinfo{person}{Sergio Gil-L{\'o}pez}, \bibinfo{person}{Daniel Molina},
  \bibinfo{person}{Richard Benjamins}, {et~al\mbox{.}}}
  \bibinfo{year}{2020}\natexlab{}.
\newblock \showarticletitle{Explainable Artificial Intelligence (XAI):
  Concepts, Taxonomies, Opportunities and Challenges toward Responsible AI}.
\newblock \bibinfo{journal}{\emph{Information Fusion}}  \bibinfo{volume}{58}
  (\bibinfo{year}{2020}), \bibinfo{pages}{82--115}.
\newblock


\bibitem[\protect\citeauthoryear{Barocas and Selbst}{Barocas and
  Selbst}{2016}]%
        {barocas2016big}
\bibfield{author}{\bibinfo{person}{Solon Barocas} {and}
  \bibinfo{person}{Andrew~D Selbst}.} \bibinfo{year}{2016}\natexlab{}.
\newblock \showarticletitle{Big Data's Disparate Impact}.
\newblock \bibinfo{journal}{\emph{Calif. L. Rev.}}  \bibinfo{volume}{104}
  (\bibinfo{year}{2016}), \bibinfo{pages}{671}.
\newblock


\bibitem[\protect\citeauthoryear{Bender, Gebru, McMillan-Major, and
  Shmitchell}{Bender et~al\mbox{.}}{2021}]%
        {bender2021dangers}
\bibfield{author}{\bibinfo{person}{Emily~M Bender}, \bibinfo{person}{Timnit
  Gebru}, \bibinfo{person}{Angelina McMillan-Major}, {and}
  \bibinfo{person}{Shmargaret Shmitchell}.} \bibinfo{year}{2021}\natexlab{}.
\newblock \showarticletitle{On the Dangers of Stochastic Parrots: Can Language
  Models Be Too Big?}. In \bibinfo{booktitle}{\emph{Proceedings of the 2021 ACM
  Conference on Fairness, Accountability, and Transparency}}.
  \bibinfo{pages}{610--623}.
\newblock


\bibitem[\protect\citeauthoryear{Benjamin}{Benjamin}{2019}]%
        {Benjamin2019AssessingRA}
\bibfield{author}{\bibinfo{person}{Ruha Benjamin}.}
  \bibinfo{year}{2019}\natexlab{}.
\newblock \showarticletitle{Assessing risk, automating racism}.
\newblock \bibinfo{journal}{\emph{Science}}  \bibinfo{volume}{366}
  (\bibinfo{year}{2019}), \bibinfo{pages}{421 -- 422}.
\newblock


\bibitem[\protect\citeauthoryear{Benthall and Haynes}{Benthall and
  Haynes}{2019}]%
        {benthall2019racial}
\bibfield{author}{\bibinfo{person}{Sebastian Benthall} {and}
  \bibinfo{person}{Bruce~D Haynes}.} \bibinfo{year}{2019}\natexlab{}.
\newblock \showarticletitle{Racial Categories in Machine Learning}. In
  \bibinfo{booktitle}{\emph{Proceedings of the conference on fairness,
  accountability, and transparency}}. \bibinfo{pages}{289--298}.
\newblock


\bibitem[\protect\citeauthoryear{Bhatt, Xiang, Sharma, Weller, Taly, Jia,
  Ghosh, Puri, Moura, and Eckersley}{Bhatt et~al\mbox{.}}{2020}]%
        {bhatt10.1145/3351095.3375624}
\bibfield{author}{\bibinfo{person}{Umang Bhatt}, \bibinfo{person}{Alice Xiang},
  \bibinfo{person}{Shubham Sharma}, \bibinfo{person}{Adrian Weller},
  \bibinfo{person}{Ankur Taly}, \bibinfo{person}{Yunhan Jia},
  \bibinfo{person}{Joydeep Ghosh}, \bibinfo{person}{Ruchir Puri},
  \bibinfo{person}{Jos\'{e} M.~F. Moura}, {and} \bibinfo{person}{Peter
  Eckersley}.} \bibinfo{year}{2020}\natexlab{}.
\newblock \showarticletitle{Explainable Machine Learning in Deployment}. In
  \bibinfo{booktitle}{\emph{Proceedings of the 2020 Conference on Fairness,
  Accountability, and Transparency}} (Barcelona, Spain)
  \emph{(\bibinfo{series}{FAT* '20})}. \bibinfo{publisher}{Association for
  Computing Machinery}, \bibinfo{address}{New York, NY, USA},
  \bibinfo{pages}{648–657}.
\newblock
\showISBNx{9781450369367}
\urldef\tempurl%
\url{https://doi.org/10.1145/3351095.3375624}
\showDOI{\tempurl}


\bibitem[\protect\citeauthoryear{Bird, Klein, and Loper}{Bird
  et~al\mbox{.}}{2009}]%
        {bird:nltk:2009}
\bibfield{author}{\bibinfo{person}{Steven Bird}, \bibinfo{person}{Ewan Klein},
  {and} \bibinfo{person}{Edward Loper}.} \bibinfo{year}{2009}\natexlab{}.
\newblock \bibinfo{booktitle}{\emph{Natural language processing with Python:
  analyzing text with the natural language toolkit}}.
\newblock \bibinfo{publisher}{" O'Reilly Media, Inc."}.
\newblock


\bibitem[\protect\citeauthoryear{Bommasani, Hudson, Adeli, Altman, Arora, von
  Arx, Bernstein, Bohg, Bosselut, Brunskill, et~al\mbox{.}}{Bommasani
  et~al\mbox{.}}{2021}]%
        {bommasani2021opportunities}
\bibfield{author}{\bibinfo{person}{Rishi Bommasani}, \bibinfo{person}{Drew~A
  Hudson}, \bibinfo{person}{Ehsan Adeli}, \bibinfo{person}{Russ Altman},
  \bibinfo{person}{Simran Arora}, \bibinfo{person}{Sydney von Arx},
  \bibinfo{person}{Michael~S Bernstein}, \bibinfo{person}{Jeannette Bohg},
  \bibinfo{person}{Antoine Bosselut}, \bibinfo{person}{Emma Brunskill},
  {et~al\mbox{.}}} \bibinfo{year}{2021}\natexlab{}.
\newblock \showarticletitle{On the Opportunities and Risks of Foundation
  Models}.
\newblock \bibinfo{journal}{\emph{arXiv preprint arXiv:2108.07258}}
  (\bibinfo{year}{2021}).
\newblock


\bibitem[\protect\citeauthoryear{Buolamwini and Gebru}{Buolamwini and
  Gebru}{2018}]%
        {buolamwini:gender_shades:2018}
\bibfield{author}{\bibinfo{person}{Joy Buolamwini} {and}
  \bibinfo{person}{Timnit Gebru}.} \bibinfo{year}{2018}\natexlab{}.
\newblock \showarticletitle{Gender Shades: Intersectional Accuracy Disparities
  in Commercial Gender Classification}. In
  \bibinfo{booktitle}{\emph{Proceedings of the 1st Conference on Fairness,
  Accountability and Transparency}} \emph{(\bibinfo{series}{Proceedings of
  Machine Learning Research}, Vol.~\bibinfo{volume}{81})},
  \bibfield{editor}{\bibinfo{person}{Sorelle~A. Friedler} {and}
  \bibinfo{person}{Christo Wilson}} (Eds.). \bibinfo{publisher}{PMLR},
  \bibinfo{pages}{77--91}.
\newblock
\urldef\tempurl%
\url{https://proceedings.mlr.press/v81/buolamwini18a.html}
\showURL{%
\tempurl}


\bibitem[\protect\citeauthoryear{Carroll, Garba, Figueroa-Rodr{\'\i}guez,
  Holbrook, Lovett, Materechera, Parsons, Raseroka, Rodriguez-Lonebear, Rowe,
  et~al\mbox{.}}{Carroll et~al\mbox{.}}{2020}]%
        {carroll2020care}
\bibfield{author}{\bibinfo{person}{Stephanie~Russo Carroll},
  \bibinfo{person}{Ibrahim Garba}, \bibinfo{person}{Oscar~L
  Figueroa-Rodr{\'\i}guez}, \bibinfo{person}{Jarita Holbrook},
  \bibinfo{person}{Raymond Lovett}, \bibinfo{person}{Simeon Materechera},
  \bibinfo{person}{Mark Parsons}, \bibinfo{person}{Kay Raseroka},
  \bibinfo{person}{Desi Rodriguez-Lonebear}, \bibinfo{person}{Robyn Rowe},
  {et~al\mbox{.}}} \bibinfo{year}{2020}\natexlab{}.
\newblock \showarticletitle{The CARE Principles for Indigenous Data
  Governance.}
\newblock  (\bibinfo{year}{2020}).
\newblock


\bibitem[\protect\citeauthoryear{Charmaz}{Charmaz}{2014}]%
        {charmaz2014constructing}
\bibfield{author}{\bibinfo{person}{Kathy Charmaz}.}
  \bibinfo{year}{2014}\natexlab{}.
\newblock \bibinfo{booktitle}{\emph{Constructing Grounded Theory}}.
\newblock \bibinfo{publisher}{sage}.
\newblock


\bibitem[\protect\citeauthoryear{Corbett-Davies and Goel}{Corbett-Davies and
  Goel}{2018}]%
        {corbett2018measure}
\bibfield{author}{\bibinfo{person}{Sam Corbett-Davies} {and}
  \bibinfo{person}{Sharad Goel}.} \bibinfo{year}{2018}\natexlab{}.
\newblock \showarticletitle{The Measure and Mismeasure of Fairness: A Critical
  Review of Fair Machine Learning}.
\newblock \bibinfo{journal}{\emph{arXiv preprint arXiv:1808.00023}}
  (\bibinfo{year}{2018}).
\newblock


\bibitem[\protect\citeauthoryear{Crisan and Fiore-Gartland}{Crisan and
  Fiore-Gartland}{2021}]%
        {Crisan:fits_and_starts:2021}
\bibfield{author}{\bibinfo{person}{Anamaria Crisan} {and}
  \bibinfo{person}{Brittany Fiore-Gartland}.} \bibinfo{year}{2021}\natexlab{}.
\newblock \bibinfo{booktitle}{\emph{Fits and Starts: Enterprise Use of AutoML
  and the Role of Humans in the Loop}}.
\newblock \bibinfo{publisher}{Association for Computing Machinery},
  \bibinfo{address}{New York, NY, USA}.
\newblock
\showISBNx{9781450380966}
\urldef\tempurl%
\url{https://doi.org/10.1145/3411764.3445775}
\showURL{%
\tempurl}


\bibitem[\protect\citeauthoryear{Czarnowska, Vyas, and Shah}{Czarnowska
  et~al\mbox{.}}{2021}]%
        {czarnowska2021quantifying}
\bibfield{author}{\bibinfo{person}{Paula Czarnowska}, \bibinfo{person}{Yogarshi
  Vyas}, {and} \bibinfo{person}{Kashif Shah}.} \bibinfo{year}{2021}\natexlab{}.
\newblock \bibinfo{title}{Quantifying Social Biases in NLP: A Generalization
  and Empirical Comparison of Extrinsic Fairness Metrics}.
\newblock
\newblock
\showeprint[arxiv]{2106.14574}~[cs.CL]


\bibitem[\protect\citeauthoryear{Dietvorst, Simmons, and Massey}{Dietvorst
  et~al\mbox{.}}{2015}]%
        {Dietvorst:AlgorithmAP:2015}
\bibfield{author}{\bibinfo{person}{Berkeley~J. Dietvorst},
  \bibinfo{person}{Joseph~P. Simmons}, {and} \bibinfo{person}{Cade Massey}.}
  \bibinfo{year}{2015}\natexlab{}.
\newblock \showarticletitle{Algorithm aversion: people erroneously avoid
  algorithms after seeing them err.}
\newblock \bibinfo{journal}{\emph{Journal of experimental psychology. General}}
   \bibinfo{volume}{144 1} (\bibinfo{year}{2015}), \bibinfo{pages}{114--26}.
\newblock


\bibitem[\protect\citeauthoryear{Dombrowski, Harmon, and Fox}{Dombrowski
  et~al\mbox{.}}{2016}]%
        {dombrowski2016social}
\bibfield{author}{\bibinfo{person}{Lynn Dombrowski}, \bibinfo{person}{Ellie
  Harmon}, {and} \bibinfo{person}{Sarah Fox}.} \bibinfo{year}{2016}\natexlab{}.
\newblock \showarticletitle{Social Justice-oriented Interaction Design:
  Outlining Key Design Strategies and Commitments}. In
  \bibinfo{booktitle}{\emph{Proceedings of the 2016 ACM Conference on Designing
  Interactive Systems}}. \bibinfo{pages}{656--671}.
\newblock


\bibitem[\protect\citeauthoryear{Dudley and Kristensson}{Dudley and
  Kristensson}{2018}]%
        {dudley:IML:2018}
\bibfield{author}{\bibinfo{person}{John~J. Dudley} {and}
  \bibinfo{person}{Per~Ola Kristensson}.} \bibinfo{year}{2018}\natexlab{}.
\newblock \showarticletitle{A Review of User Interface Design for Interactive
  Machine Learning}.
\newblock \bibinfo{journal}{\emph{ACM Trans. Interact. Intell. Syst.}}
  \bibinfo{volume}{8}, \bibinfo{number}{2}, Article \bibinfo{articleno}{8}
  (\bibinfo{date}{jun} \bibinfo{year}{2018}), \bibinfo{numpages}{37}~pages.
\newblock
\showISSN{2160-6455}
\urldef\tempurl%
\url{https://doi.org/10.1145/3185517}
\showDOI{\tempurl}


\bibitem[\protect\citeauthoryear{Ehsan and Riedl}{Ehsan and Riedl}{2021}]%
        {ehsan2021explainability}
\bibfield{author}{\bibinfo{person}{Upol Ehsan} {and} \bibinfo{person}{Mark~O
  Riedl}.} \bibinfo{year}{2021}\natexlab{}.
\newblock \showarticletitle{Explainability Pitfalls: Beyond Dark Patterns in
  Explainable AI}.
\newblock \bibinfo{journal}{\emph{arXiv preprint arXiv:2109.12480}}
  (\bibinfo{year}{2021}).
\newblock


\bibitem[\protect\citeauthoryear{Gebru, Morgenstern, Vecchione, Vaughan,
  Wallach, Iii, and Crawford}{Gebru et~al\mbox{.}}{2021}]%
        {gebru2021datasheets}
\bibfield{author}{\bibinfo{person}{Timnit Gebru}, \bibinfo{person}{Jamie
  Morgenstern}, \bibinfo{person}{Briana Vecchione},
  \bibinfo{person}{Jennifer~Wortman Vaughan}, \bibinfo{person}{Hanna Wallach},
  \bibinfo{person}{Hal~Daum{\'e} Iii}, {and} \bibinfo{person}{Kate Crawford}.}
  \bibinfo{year}{2021}\natexlab{}.
\newblock \showarticletitle{Datasheets for Datasets}.
\newblock \bibinfo{journal}{\emph{Commun. ACM}} \bibinfo{volume}{64},
  \bibinfo{number}{12} (\bibinfo{year}{2021}), \bibinfo{pages}{86--92}.
\newblock


\bibitem[\protect\citeauthoryear{Giannopoulou}{Giannopoulou}{2020}]%
        {giannopoulou2020algorithmic}
\bibfield{author}{\bibinfo{person}{Alexandra Giannopoulou}.}
  \bibinfo{year}{2020}\natexlab{}.
\newblock \showarticletitle{Algorithmic Systems: the Consent is in the Detail?}
\newblock \bibinfo{journal}{\emph{Internet Policy Review}} \bibinfo{volume}{9},
  \bibinfo{number}{1} (\bibinfo{year}{2020}).
\newblock


\bibitem[\protect\citeauthoryear{Goel, Rajani, Vig, Tan, Wu, Zheng, Xiong,
  Bansal, and R{\'e}}{Goel et~al\mbox{.}}{2021}]%
        {goel2021robustness}
\bibfield{author}{\bibinfo{person}{Karan Goel}, \bibinfo{person}{Nazneen
  Rajani}, \bibinfo{person}{Jesse Vig}, \bibinfo{person}{Samson Tan},
  \bibinfo{person}{Jason Wu}, \bibinfo{person}{Stephan Zheng},
  \bibinfo{person}{Caiming Xiong}, \bibinfo{person}{Mohit Bansal}, {and}
  \bibinfo{person}{Christopher R{\'e}}.} \bibinfo{year}{2021}\natexlab{}.
\newblock \showarticletitle{Robustness Gym: Unifying the NLP Evaluation
  Landscape}.
\newblock \bibinfo{journal}{\emph{arXiv preprint arXiv:2101.04840}}
  (\bibinfo{year}{2021}).
\newblock


\bibitem[\protect\citeauthoryear{Guidotti, Monreale, Ruggieri, Turini,
  Giannotti, and Pedreschi}{Guidotti et~al\mbox{.}}{2018}]%
        {guidotti2018survey}
\bibfield{author}{\bibinfo{person}{Riccardo Guidotti}, \bibinfo{person}{Anna
  Monreale}, \bibinfo{person}{Salvatore Ruggieri}, \bibinfo{person}{Franco
  Turini}, \bibinfo{person}{Fosca Giannotti}, {and} \bibinfo{person}{Dino
  Pedreschi}.} \bibinfo{year}{2018}\natexlab{}.
\newblock \showarticletitle{A Survey of Methods for Explaining Black Box
  Models}.
\newblock \bibinfo{journal}{\emph{ACM computing surveys (CSUR)}}
  \bibinfo{volume}{51}, \bibinfo{number}{5} (\bibinfo{year}{2018}),
  \bibinfo{pages}{1--42}.
\newblock


\bibitem[\protect\citeauthoryear{Hamidi, Scheuerman, and Branham}{Hamidi
  et~al\mbox{.}}{2018}]%
        {hamidi2018gender}
\bibfield{author}{\bibinfo{person}{Foad Hamidi}, \bibinfo{person}{Morgan~Klaus
  Scheuerman}, {and} \bibinfo{person}{Stacy~M Branham}.}
  \bibinfo{year}{2018}\natexlab{}.
\newblock \showarticletitle{Gender Recognition or Gender Reductionism? The
  Social Implications of Embedded Gender Recognition Systems}. In
  \bibinfo{booktitle}{\emph{Proceedings of the 2018 chi conference on human
  factors in computing systems}}. \bibinfo{pages}{1--13}.
\newblock


\bibitem[\protect\citeauthoryear{Hanna, Denton, Smart, and Smith-Loud}{Hanna
  et~al\mbox{.}}{2020}]%
        {hanna2020towards}
\bibfield{author}{\bibinfo{person}{Alex Hanna}, \bibinfo{person}{Emily Denton},
  \bibinfo{person}{Andrew Smart}, {and} \bibinfo{person}{Jamila Smith-Loud}.}
  \bibinfo{year}{2020}\natexlab{}.
\newblock \showarticletitle{Towards a Critical Race Methodology in Algorithmic
  Fairness}. In \bibinfo{booktitle}{\emph{Proceedings of the 2020 conference on
  fairness, accountability, and transparency}}. \bibinfo{pages}{501--512}.
\newblock


\bibitem[\protect\citeauthoryear{Harris, Millman, van~der Walt, Gommers,
  Virtanen, Cournapeau, Wieser, Taylor, Berg, Smith, Kern, Picus, Hoyer, van
  Kerkwijk, Brett, Haldane, Fernández~del Río, Wiebe, Peterson,
  Gérard-Marchant, Sheppard, Reddy, Weckesser, Abbasi, Gohlke, and
  Oliphant}{Harris et~al\mbox{.}}{2020}]%
        {harris:numpy:2020}
\bibfield{author}{\bibinfo{person}{Charles~R. Harris},
  \bibinfo{person}{K.~Jarrod Millman}, \bibinfo{person}{Stéfan~J van~der
  Walt}, \bibinfo{person}{Ralf Gommers}, \bibinfo{person}{Pauli Virtanen},
  \bibinfo{person}{David Cournapeau}, \bibinfo{person}{Eric Wieser},
  \bibinfo{person}{Julian Taylor}, \bibinfo{person}{Sebastian Berg},
  \bibinfo{person}{Nathaniel~J. Smith}, \bibinfo{person}{Robert Kern},
  \bibinfo{person}{Matti Picus}, \bibinfo{person}{Stephan Hoyer},
  \bibinfo{person}{Marten~H. van Kerkwijk}, \bibinfo{person}{Matthew Brett},
  \bibinfo{person}{Allan Haldane}, \bibinfo{person}{Jaime Fernández~del Río},
  \bibinfo{person}{Mark Wiebe}, \bibinfo{person}{Pearu Peterson},
  \bibinfo{person}{Pierre Gérard-Marchant}, \bibinfo{person}{Kevin Sheppard},
  \bibinfo{person}{Tyler Reddy}, \bibinfo{person}{Warren Weckesser},
  \bibinfo{person}{Hameer Abbasi}, \bibinfo{person}{Christoph Gohlke}, {and}
  \bibinfo{person}{Travis~E. Oliphant}.} \bibinfo{year}{2020}\natexlab{}.
\newblock \showarticletitle{Array programming with {NumPy}}.
\newblock \bibinfo{journal}{\emph{Nature}}  \bibinfo{volume}{585}
  (\bibinfo{year}{2020}), \bibinfo{pages}{357–362}.
\newblock
\urldef\tempurl%
\url{https://doi.org/10.1038/s41586-020-2649-2}
\showDOI{\tempurl}


\bibitem[\protect\citeauthoryear{Hirsch, Merced, Narayanan, Imel, and
  Atkins}{Hirsch et~al\mbox{.}}{2017}]%
        {hirsch2017designing}
\bibfield{author}{\bibinfo{person}{Tad Hirsch}, \bibinfo{person}{Kritzia
  Merced}, \bibinfo{person}{Shrikanth Narayanan}, \bibinfo{person}{Zac~E Imel},
  {and} \bibinfo{person}{David~C Atkins}.} \bibinfo{year}{2017}\natexlab{}.
\newblock \showarticletitle{Designing Contestability: Interaction Design,
  Machine Learning, and Mental Health}. In
  \bibinfo{booktitle}{\emph{Proceedings of the 2017 Conference on Designing
  Interactive Systems}}. \bibinfo{pages}{95--99}.
\newblock


\bibitem[\protect\citeauthoryear{Hoffmann}{Hoffmann}{2019}]%
        {hoffmann2019fairness}
\bibfield{author}{\bibinfo{person}{Anna~Lauren Hoffmann}.}
  \bibinfo{year}{2019}\natexlab{}.
\newblock \showarticletitle{Where Fairness Fails: Data, Algorithms, and the
  Limits of Antidiscrimination Discourse}.
\newblock \bibinfo{journal}{\emph{Information, Communication \& Society}}
  \bibinfo{volume}{22}, \bibinfo{number}{7} (\bibinfo{year}{2019}),
  \bibinfo{pages}{900--915}.
\newblock


\bibitem[\protect\citeauthoryear{Hohman, Conlen, Heer, and Chau}{Hohman
  et~al\mbox{.}}{2020}]%
        {hohman:explainable:2020}
\bibfield{author}{\bibinfo{person}{Fred Hohman}, \bibinfo{person}{Matthew
  Conlen}, \bibinfo{person}{Jeffrey Heer}, {and} \bibinfo{person}{Duen
  Horng~(Polo) Chau}.} \bibinfo{year}{2020}\natexlab{}.
\newblock \showarticletitle{Communicating with Interactive Articles}.
\newblock \bibinfo{journal}{\emph{Distill}} (\bibinfo{year}{2020}).
\newblock
\urldef\tempurl%
\url{https://doi.org/10.23915/distill.00028}
\showDOI{\tempurl}
\newblock
\shownote{https://distill.pub/2020/communicating-with-interactive-articles.}


\bibitem[\protect\citeauthoryear{Holstein, Wortman~Vaughan, Daum{\'e}~III,
  Dudik, and Wallach}{Holstein et~al\mbox{.}}{2019}]%
        {holstein2019improving}
\bibfield{author}{\bibinfo{person}{Kenneth Holstein}, \bibinfo{person}{Jennifer
  Wortman~Vaughan}, \bibinfo{person}{Hal Daum{\'e}~III}, \bibinfo{person}{Miro
  Dudik}, {and} \bibinfo{person}{Hanna Wallach}.}
  \bibinfo{year}{2019}\natexlab{}.
\newblock \showarticletitle{Improving Fairness in Machine Learning Systems:
  What do Industry Practitioners Need?}. In
  \bibinfo{booktitle}{\emph{Proceedings of the 2019 CHI conference on human
  factors in computing systems}}. \bibinfo{pages}{1--16}.
\newblock


\bibitem[\protect\citeauthoryear{Hong, Hullman, and Bertini}{Hong
  et~al\mbox{.}}{2020}]%
        {hong2020human}
\bibfield{author}{\bibinfo{person}{Sungsoo~Ray Hong}, \bibinfo{person}{Jessica
  Hullman}, {and} \bibinfo{person}{Enrico Bertini}.}
  \bibinfo{year}{2020}\natexlab{}.
\newblock \showarticletitle{Human Factors in Model Interpretability: Industry
  Practices, Challenges, and Needs}.
\newblock \bibinfo{journal}{\emph{Proceedings of the ACM on Human-Computer
  Interaction}} \bibinfo{volume}{4}, \bibinfo{number}{CSCW1}
  (\bibinfo{year}{2020}), \bibinfo{pages}{1--26}.
\newblock


\bibitem[\protect\citeauthoryear{Jacovi, Marasovi{\'c}, Miller, and
  Goldberg}{Jacovi et~al\mbox{.}}{2021}]%
        {jacovi2021formalizing}
\bibfield{author}{\bibinfo{person}{Alon Jacovi}, \bibinfo{person}{Ana
  Marasovi{\'c}}, \bibinfo{person}{Tim Miller}, {and} \bibinfo{person}{Yoav
  Goldberg}.} \bibinfo{year}{2021}\natexlab{}.
\newblock \showarticletitle{Formalizing Trust in Artificial Intelligence:
  Prerequisites, Causes and Goals of Human Trust in AI}. In
  \bibinfo{booktitle}{\emph{Proceedings of the 2021 ACM Conference on Fairness,
  Accountability, and Transparency}}. \bibinfo{pages}{624--635}.
\newblock


\bibitem[\protect\citeauthoryear{Kaur, Nori, Jenkins, Caruana, Wallach, and
  Wortman~Vaughan}{Kaur et~al\mbox{.}}{2020}]%
        {kaur2020interpreting}
\bibfield{author}{\bibinfo{person}{Harmanpreet Kaur}, \bibinfo{person}{Harsha
  Nori}, \bibinfo{person}{Samuel Jenkins}, \bibinfo{person}{Rich Caruana},
  \bibinfo{person}{Hanna Wallach}, {and} \bibinfo{person}{Jennifer
  Wortman~Vaughan}.} \bibinfo{year}{2020}\natexlab{}.
\newblock \showarticletitle{Interpreting Interpretability: Understanding Data
  Scientists' Use of Interpretability Tools for Machine Learning}. In
  \bibinfo{booktitle}{\emph{Proceedings of the 2020 CHI Conference on Human
  Factors in Computing Systems}}. \bibinfo{pages}{1--14}.
\newblock


\bibitem[\protect\citeauthoryear{Kingsley, Wang, Mikhalenko, Sinha, and
  Kulkarni}{Kingsley et~al\mbox{.}}{2020}]%
        {kingsley2020auditing}
\bibfield{author}{\bibinfo{person}{Sara Kingsley}, \bibinfo{person}{Clara
  Wang}, \bibinfo{person}{Alex Mikhalenko}, \bibinfo{person}{Proteeti Sinha},
  {and} \bibinfo{person}{Chinmay Kulkarni}.} \bibinfo{year}{2020}\natexlab{}.
\newblock \showarticletitle{Auditing Digital Platforms for Discrimination in
  Economic Opportunity Advertising}.
\newblock \bibinfo{journal}{\emph{arXiv preprint arXiv:2008.09656}}
  (\bibinfo{year}{2020}).
\newblock


\bibitem[\protect\citeauthoryear{Knijnenburg, Reijmer, and
  Willemsen}{Knijnenburg et~al\mbox{.}}{2011}]%
        {knijnenburg2011each}
\bibfield{author}{\bibinfo{person}{Bart~P Knijnenburg},
  \bibinfo{person}{Niels~JM Reijmer}, {and} \bibinfo{person}{Martijn~C
  Willemsen}.} \bibinfo{year}{2011}\natexlab{}.
\newblock \showarticletitle{Each to His Own: How Different Users Call for
  Different Interaction Methods in Recommender Systems}. In
  \bibinfo{booktitle}{\emph{Proceedings of the fifth ACM conference on
  Recommender systems}}. \bibinfo{pages}{141--148}.
\newblock


\bibitem[\protect\citeauthoryear{Krafft, Young, Katell, Lee, Narayan, Epstein,
  Dailey, Herman, Tam, Guetler, Bintz, Raz, Jobe, Putz, Robick, and
  Barghouti}{Krafft et~al\mbox{.}}{2021}]%
        {krafft2021Action}
\bibfield{author}{\bibinfo{person}{P.~M. Krafft}, \bibinfo{person}{Meg Young},
  \bibinfo{person}{Michael Katell}, \bibinfo{person}{Jennifer~E. Lee},
  \bibinfo{person}{Shankar Narayan}, \bibinfo{person}{Micah Epstein},
  \bibinfo{person}{Dharma Dailey}, \bibinfo{person}{Bernease Herman},
  \bibinfo{person}{Aaron Tam}, \bibinfo{person}{Vivian Guetler},
  \bibinfo{person}{Corinne Bintz}, \bibinfo{person}{Daniella Raz},
  \bibinfo{person}{Pa~Ousman Jobe}, \bibinfo{person}{Franziska Putz},
  \bibinfo{person}{Brian Robick}, {and} \bibinfo{person}{Bissan Barghouti}.}
  \bibinfo{year}{2021}\natexlab{}.
\newblock \showarticletitle{An Action-Oriented AI Policy Toolkit for Technology
  Audits by Community Advocates and Activists}. In
  \bibinfo{booktitle}{\emph{Proceedings of the 2021 ACM Conference on Fairness,
  Accountability, and Transparency}} (Virtual Event, Canada)
  \emph{(\bibinfo{series}{FAccT '21})}. \bibinfo{publisher}{Association for
  Computing Machinery}, \bibinfo{address}{New York, NY, USA},
  \bibinfo{pages}{772–781}.
\newblock
\showISBNx{9781450383097}
\urldef\tempurl%
\url{https://doi.org/10.1145/3442188.3445938}
\showDOI{\tempurl}


\bibitem[\protect\citeauthoryear{Lee, Kusbit, Kahng, Kim, Yuan, Chan, See,
  Noothigattu, Lee, Psomas, et~al\mbox{.}}{Lee et~al\mbox{.}}{2019}]%
        {lee2019webuildai}
\bibfield{author}{\bibinfo{person}{Min~Kyung Lee}, \bibinfo{person}{Daniel
  Kusbit}, \bibinfo{person}{Anson Kahng}, \bibinfo{person}{Ji~Tae Kim},
  \bibinfo{person}{Xinran Yuan}, \bibinfo{person}{Allissa Chan},
  \bibinfo{person}{Daniel See}, \bibinfo{person}{Ritesh Noothigattu},
  \bibinfo{person}{Siheon Lee}, \bibinfo{person}{Alexandros Psomas},
  {et~al\mbox{.}}} \bibinfo{year}{2019}\natexlab{}.
\newblock \showarticletitle{WeBuildAI: Participatory Framework for Algorithmic
  Governance}.
\newblock \bibinfo{journal}{\emph{Proceedings of the ACM on Human-Computer
  Interaction}} \bibinfo{volume}{3}, \bibinfo{number}{CSCW}
  (\bibinfo{year}{2019}), \bibinfo{pages}{1--35}.
\newblock


\bibitem[\protect\citeauthoryear{Lucic, Haned, and de~Rijke}{Lucic
  et~al\mbox{.}}{2020}]%
        {lucic10.1145/3351095.3372824}
\bibfield{author}{\bibinfo{person}{Ana Lucic}, \bibinfo{person}{Hinda Haned},
  {and} \bibinfo{person}{Maarten de Rijke}.} \bibinfo{year}{2020}\natexlab{}.
\newblock \showarticletitle{Why Does My Model Fail? Contrastive Local
  Explanations for Retail Forecasting}. In
  \bibinfo{booktitle}{\emph{Proceedings of the 2020 Conference on Fairness,
  Accountability, and Transparency}} (Barcelona, Spain)
  \emph{(\bibinfo{series}{FAT* '20})}. \bibinfo{publisher}{Association for
  Computing Machinery}, \bibinfo{address}{New York, NY, USA},
  \bibinfo{pages}{90–98}.
\newblock
\showISBNx{9781450369367}
\urldef\tempurl%
\url{https://doi.org/10.1145/3351095.3372824}
\showDOI{\tempurl}


\bibitem[\protect\citeauthoryear{Lyons, Velloso, and Miller}{Lyons
  et~al\mbox{.}}{2021}]%
        {lyons10.1145/3449180}
\bibfield{author}{\bibinfo{person}{Henrietta Lyons}, \bibinfo{person}{Eduardo
  Velloso}, {and} \bibinfo{person}{Tim Miller}.}
  \bibinfo{year}{2021}\natexlab{}.
\newblock \showarticletitle{Conceptualising Contestability: Perspectives on
  Contesting Algorithmic Decisions}.
\newblock \bibinfo{journal}{\emph{Proc. ACM Hum.-Comput. Interact.}}
  \bibinfo{volume}{5}, \bibinfo{number}{CSCW1}, Article
  \bibinfo{articleno}{106} (\bibinfo{date}{apr} \bibinfo{year}{2021}),
  \bibinfo{numpages}{25}~pages.
\newblock
\urldef\tempurl%
\url{https://doi.org/10.1145/3449180}
\showDOI{\tempurl}


\bibitem[\protect\citeauthoryear{Madsen, Reddy, and Chandar}{Madsen
  et~al\mbox{.}}{2021}]%
        {madsen2021post}
\bibfield{author}{\bibinfo{person}{Andreas Madsen}, \bibinfo{person}{Siva
  Reddy}, {and} \bibinfo{person}{Sarath Chandar}.}
  \bibinfo{year}{2021}\natexlab{}.
\newblock \showarticletitle{Post-hoc Interpretability for Neural NLP: A
  Survey}.
\newblock \bibinfo{journal}{\emph{arXiv preprint arXiv:2108.04840}}
  (\bibinfo{year}{2021}).
\newblock


\bibitem[\protect\citeauthoryear{McKinney et~al\mbox{.}}{McKinney
  et~al\mbox{.}}{2010}]%
        {mckinney:pandas:2010}
\bibfield{author}{\bibinfo{person}{Wes McKinney} {et~al\mbox{.}}}
  \bibinfo{year}{2010}\natexlab{}.
\newblock \showarticletitle{Data structures for statistical computing in
  python}. In \bibinfo{booktitle}{\emph{Proceedings of the 9th Python in
  Science Conference}}, Vol.~\bibinfo{volume}{445}. Austin, TX,
  \bibinfo{pages}{51--56}.
\newblock


\bibitem[\protect\citeauthoryear{Mishra and Rzeszotarski}{Mishra and
  Rzeszotarski}{2021}]%
        {mishra2021designing}
\bibfield{author}{\bibinfo{person}{Swati Mishra} {and}
  \bibinfo{person}{Jeffrey~M Rzeszotarski}.} \bibinfo{year}{2021}\natexlab{}.
\newblock \showarticletitle{Designing Interactive Transfer Learning Tools for
  ML Non-Experts}. In \bibinfo{booktitle}{\emph{Proceedings of the 2021 CHI
  Conference on Human Factors in Computing Systems}}. \bibinfo{pages}{1--15}.
\newblock


\bibitem[\protect\citeauthoryear{Mitchell, Wu, Zaldivar, Barnes, Vasserman,
  Hutchinson, Spitzer, Raji, and Gebru}{Mitchell et~al\mbox{.}}{2019}]%
        {mitchell10.1145/3287560.3287596}
\bibfield{author}{\bibinfo{person}{Margaret Mitchell}, \bibinfo{person}{Simone
  Wu}, \bibinfo{person}{Andrew Zaldivar}, \bibinfo{person}{Parker Barnes},
  \bibinfo{person}{Lucy Vasserman}, \bibinfo{person}{Ben Hutchinson},
  \bibinfo{person}{Elena Spitzer}, \bibinfo{person}{Inioluwa~Deborah Raji},
  {and} \bibinfo{person}{Timnit Gebru}.} \bibinfo{year}{2019}\natexlab{}.
\newblock \showarticletitle{Model Cards for Model Reporting}. In
  \bibinfo{booktitle}{\emph{Proceedings of the Conference on Fairness,
  Accountability, and Transparency}} (Atlanta, GA, USA)
  \emph{(\bibinfo{series}{FAT* '19})}. \bibinfo{publisher}{Association for
  Computing Machinery}, \bibinfo{address}{New York, NY, USA},
  \bibinfo{pages}{220–229}.
\newblock
\showISBNx{9781450361255}
\urldef\tempurl%
\url{https://doi.org/10.1145/3287560.3287596}
\showDOI{\tempurl}


\bibitem[\protect\citeauthoryear{Mittelstadt, Russell, and Wachter}{Mittelstadt
  et~al\mbox{.}}{2019}]%
        {mittelstadt10.1145/3287560.3287574}
\bibfield{author}{\bibinfo{person}{Brent Mittelstadt}, \bibinfo{person}{Chris
  Russell}, {and} \bibinfo{person}{Sandra Wachter}.}
  \bibinfo{year}{2019}\natexlab{}.
\newblock \showarticletitle{Explaining Explanations in AI}. In
  \bibinfo{booktitle}{\emph{Proceedings of the Conference on Fairness,
  Accountability, and Transparency}} (Atlanta, GA, USA)
  \emph{(\bibinfo{series}{FAT* '19})}. \bibinfo{publisher}{Association for
  Computing Machinery}, \bibinfo{address}{New York, NY, USA},
  \bibinfo{pages}{279–288}.
\newblock
\showISBNx{9781450361255}
\urldef\tempurl%
\url{https://doi.org/10.1145/3287560.3287574}
\showDOI{\tempurl}


\bibitem[\protect\citeauthoryear{Mokander and Floridi}{Mokander and
  Floridi}{2021}]%
        {mokander2021ethics}
\bibfield{author}{\bibinfo{person}{Jakob Mokander} {and}
  \bibinfo{person}{Luciano Floridi}.} \bibinfo{year}{2021}\natexlab{}.
\newblock \showarticletitle{Ethics-based Auditing to Develop Trustworthy AI}.
\newblock \bibinfo{journal}{\emph{arXiv preprint arXiv:2105.00002}}
  (\bibinfo{year}{2021}).
\newblock


\bibitem[\protect\citeauthoryear{Molnar, Casalicchio, and Bischl}{Molnar
  et~al\mbox{.}}{2020}]%
        {molnar2020interpretable}
\bibfield{author}{\bibinfo{person}{Christoph Molnar}, \bibinfo{person}{Giuseppe
  Casalicchio}, {and} \bibinfo{person}{Bernd Bischl}.}
  \bibinfo{year}{2020}\natexlab{}.
\newblock \showarticletitle{Interpretable Machine Learning--A Brief History,
  State-of-the-Art and Challenges}. In \bibinfo{booktitle}{\emph{Joint European
  Conference on Machine Learning and Knowledge Discovery in Databases}}.
  Springer, \bibinfo{pages}{417--431}.
\newblock


\bibitem[\protect\citeauthoryear{Morley, Floridi, Kinsey, and Elhalal}{Morley
  et~al\mbox{.}}{2019}]%
        {morley:what_to_how:2019}
\bibfield{author}{\bibinfo{person}{Jessica Morley}, \bibinfo{person}{Luciano
  Floridi}, \bibinfo{person}{Libby Kinsey}, {and} \bibinfo{person}{Anat
  Elhalal}.} \bibinfo{year}{2019}\natexlab{}.
\newblock \bibinfo{title}{From What to How: An Initial Review of Publicly
  Available AI Ethics Tools, Methods and Research to Translate Principles into
  Practices}.
\newblock
\newblock
\showeprint[arxiv]{1905.06876}~[cs.CY]


\bibitem[\protect\citeauthoryear{Mozafari, Farahbakhsh, and Crespi}{Mozafari
  et~al\mbox{.}}{2020}]%
        {mozafari2020hate}
\bibfield{author}{\bibinfo{person}{Marzieh Mozafari}, \bibinfo{person}{Reza
  Farahbakhsh}, {and} \bibinfo{person}{No{\"e}l Crespi}.}
  \bibinfo{year}{2020}\natexlab{}.
\newblock \showarticletitle{Hate Speech Detection and Racial Bias Mitigation in
  Social Media Based on BERT Model}.
\newblock \bibinfo{journal}{\emph{PloS one}} \bibinfo{volume}{15},
  \bibinfo{number}{8} (\bibinfo{year}{2020}), \bibinfo{pages}{e0237861}.
\newblock


\bibitem[\protect\citeauthoryear{Muller}{Muller}{2014}]%
        {muller2014curiosity}
\bibfield{author}{\bibinfo{person}{Michael Muller}.}
  \bibinfo{year}{2014}\natexlab{}.
\newblock \showarticletitle{Curiosity, Creativity, and Surprise as Analytic
  Tools: Grounded Theory Method}.
\newblock In \bibinfo{booktitle}{\emph{Ways of Knowing in HCI}}.
  \bibinfo{publisher}{Springer}, \bibinfo{pages}{25--48}.
\newblock


\bibitem[\protect\citeauthoryear{Nourani, King, and Ragan}{Nourani
  et~al\mbox{.}}{2020}]%
        {nourani2020role}
\bibfield{author}{\bibinfo{person}{Mahsan Nourani}, \bibinfo{person}{Joanie
  King}, {and} \bibinfo{person}{Eric Ragan}.} \bibinfo{year}{2020}\natexlab{}.
\newblock \showarticletitle{The Role of Domain Expertise in User Trust and the
  Impact of First Impressions with Intelligent Systems}. In
  \bibinfo{booktitle}{\emph{Proceedings of the AAAI Conference on Human
  Computation and Crowdsourcing}}, Vol.~\bibinfo{volume}{8}.
  \bibinfo{pages}{112--121}.
\newblock


\bibitem[\protect\citeauthoryear{Nourani, Roy, Block, Honeycutt, Rahman, Ragan,
  and Gogate}{Nourani et~al\mbox{.}}{2021}]%
        {Nourani:anchoring:2021}
\bibfield{author}{\bibinfo{person}{Mahsan Nourani}, \bibinfo{person}{Chiradeep
  Roy}, \bibinfo{person}{Jeremy~E Block}, \bibinfo{person}{Donald~R Honeycutt},
  \bibinfo{person}{Tahrima Rahman}, \bibinfo{person}{Eric Ragan}, {and}
  \bibinfo{person}{Vibhav Gogate}.} \bibinfo{year}{2021}\natexlab{}.
\newblock \showarticletitle{Anchoring Bias Affects Mental Model Formation and
  User Reliance in Explainable AI Systems}. In \bibinfo{booktitle}{\emph{Proc.
  IUI'21}}. \bibinfo{pages}{340–350}.
\newblock
\urldef\tempurl%
\url{https://doi.org/10.1145/3397481.3450639}
\showURL{%
\tempurl}


\bibitem[\protect\citeauthoryear{Pedregosa, Varoquaux, Gramfort, Michel,
  Thirion, Grisel, Blondel, Prettenhofer, Weiss, Dubourg, Vanderplas, Passos,
  Cournapeau, Brucher, Perrot, and Duchesnay}{Pedregosa et~al\mbox{.}}{2011}]%
        {scikit-learn}
\bibfield{author}{\bibinfo{person}{F. Pedregosa}, \bibinfo{person}{G.
  Varoquaux}, \bibinfo{person}{A. Gramfort}, \bibinfo{person}{V. Michel},
  \bibinfo{person}{B. Thirion}, \bibinfo{person}{O. Grisel},
  \bibinfo{person}{M. Blondel}, \bibinfo{person}{P. Prettenhofer},
  \bibinfo{person}{R. Weiss}, \bibinfo{person}{V. Dubourg}, \bibinfo{person}{J.
  Vanderplas}, \bibinfo{person}{A. Passos}, \bibinfo{person}{D. Cournapeau},
  \bibinfo{person}{M. Brucher}, \bibinfo{person}{M. Perrot}, {and}
  \bibinfo{person}{E. Duchesnay}.} \bibinfo{year}{2011}\natexlab{}.
\newblock \showarticletitle{Scikit-learn: Machine Learning in {P}ython}.
\newblock \bibinfo{journal}{\emph{Journal of Machine Learning Research}}
  \bibinfo{volume}{12} (\bibinfo{year}{2011}), \bibinfo{pages}{2825--2830}.
\newblock


\bibitem[\protect\citeauthoryear{Potts, Wu, Geiger, and Kiela}{Potts
  et~al\mbox{.}}{2020}]%
        {potts2020dynasent}
\bibfield{author}{\bibinfo{person}{Christopher Potts},
  \bibinfo{person}{Zhengxuan Wu}, \bibinfo{person}{Atticus Geiger}, {and}
  \bibinfo{person}{Douwe Kiela}.} \bibinfo{year}{2020}\natexlab{}.
\newblock \showarticletitle{DynaSent: A Dynamic Benchmark for Sentiment
  Analysis}.
\newblock  (\bibinfo{year}{2020}).
\newblock
\showeprint[arxiv]{2012.15349}~[cs.CL]


\bibitem[\protect\citeauthoryear{Poursabzi-Sangdeh, Goldstein, Hofman,
  Wortman~Vaughan, and Wallach}{Poursabzi-Sangdeh et~al\mbox{.}}{2021}]%
        {poursabzi2021manipulating}
\bibfield{author}{\bibinfo{person}{Forough Poursabzi-Sangdeh},
  \bibinfo{person}{Daniel~G Goldstein}, \bibinfo{person}{Jake~M Hofman},
  \bibinfo{person}{Jennifer~Wortman Wortman~Vaughan}, {and}
  \bibinfo{person}{Hanna Wallach}.} \bibinfo{year}{2021}\natexlab{}.
\newblock \showarticletitle{Manipulating and Measuring Model Interpretability}.
  In \bibinfo{booktitle}{\emph{Proceedings of the 2021 CHI Conference on Human
  Factors in Computing Systems}}. \bibinfo{pages}{1--52}.
\newblock


\bibitem[\protect\citeauthoryear{Raji and Buolamwini}{Raji and
  Buolamwini}{2019}]%
        {raji2019actionable}
\bibfield{author}{\bibinfo{person}{Inioluwa~Deborah Raji} {and}
  \bibinfo{person}{Joy Buolamwini}.} \bibinfo{year}{2019}\natexlab{}.
\newblock \showarticletitle{Actionable Auditing: Investigating the Impact of
  Publicly Naming Biased Performance Results of Commercial AI Products}. In
  \bibinfo{booktitle}{\emph{Proceedings of the 2019 AAAI/ACM Conference on AI,
  Ethics, and Society}}. \bibinfo{pages}{429--435}.
\newblock


\bibitem[\protect\citeauthoryear{Rakova, Yang, Cramer, and Chowdhury}{Rakova
  et~al\mbox{.}}{2021}]%
        {rakova10.1145/3449081}
\bibfield{author}{\bibinfo{person}{Bogdana Rakova}, \bibinfo{person}{Jingying
  Yang}, \bibinfo{person}{Henriette Cramer}, {and} \bibinfo{person}{Rumman
  Chowdhury}.} \bibinfo{year}{2021}\natexlab{}.
\newblock \showarticletitle{Where Responsible AI Meets Reality: Practitioner
  Perspectives on Enablers for Shifting Organizational Practices}.
\newblock \bibinfo{journal}{\emph{Proc. ACM Hum.-Comput. Interact.}}
  \bibinfo{volume}{5}, \bibinfo{number}{CSCW1}, Article \bibinfo{articleno}{7}
  (\bibinfo{date}{apr} \bibinfo{year}{2021}), \bibinfo{numpages}{23}~pages.
\newblock
\urldef\tempurl%
\url{https://doi.org/10.1145/3449081}
\showDOI{\tempurl}


\bibitem[\protect\citeauthoryear{Rehurek and Sojka}{Rehurek and Sojka}{2011}]%
        {rehurek:gensim:2011}
\bibfield{author}{\bibinfo{person}{Radim Rehurek} {and} \bibinfo{person}{Petr
  Sojka}.} \bibinfo{year}{2011}\natexlab{}.
\newblock \showarticletitle{Gensim--python framework for vector space
  modelling}.
\newblock \bibinfo{journal}{\emph{NLP Centre, Faculty of Informatics, Masaryk
  University, Brno, Czech Republic}} \bibinfo{volume}{3}, \bibinfo{number}{2}
  (\bibinfo{year}{2011}).
\newblock


\bibitem[\protect\citeauthoryear{Sacha, Sedlmair, Zhang, Lee, Peltonen,
  Weiskopf, North, and Keim}{Sacha et~al\mbox{.}}{2017}]%
        {scaha:HCD:2017}
\bibfield{author}{\bibinfo{person}{Dominik Sacha}, \bibinfo{person}{Michael
  Sedlmair}, \bibinfo{person}{Leishi Zhang}, \bibinfo{person}{John~A. Lee},
  \bibinfo{person}{Jaakko Peltonen}, \bibinfo{person}{Daniel Weiskopf},
  \bibinfo{person}{Stephen~C. North}, {and} \bibinfo{person}{Daniel~A. Keim}.}
  \bibinfo{year}{2017}\natexlab{}.
\newblock \showarticletitle{What you see is what you can change: Human-centered
  machine learning by interactive visualization}.
\newblock \bibinfo{journal}{\emph{Neurocomputing}}  \bibinfo{volume}{268}
  (\bibinfo{year}{2017}), \bibinfo{pages}{164--175}.
\newblock
\showISSN{0925-2312}
\urldef\tempurl%
\url{https://doi.org/10.1016/j.neucom.2017.01.105}
\showDOI{\tempurl}
\newblock
\shownote{Advances in artificial neural networks, machine learning and
  computational intelligence.}


\bibitem[\protect\citeauthoryear{Sambasivan, Kapania, Highfill, Akrong,
  Paritosh, and Aroyo}{Sambasivan et~al\mbox{.}}{2021}]%
        {sambasivan:datawork:2021}
\bibfield{author}{\bibinfo{person}{Nithya Sambasivan}, \bibinfo{person}{Shivani
  Kapania}, \bibinfo{person}{Hannah Highfill}, \bibinfo{person}{Diana Akrong},
  \bibinfo{person}{Praveen Paritosh}, {and} \bibinfo{person}{Lora~M Aroyo}.}
  \bibinfo{year}{2021}\natexlab{}.
\newblock \bibinfo{booktitle}{\emph{“Everyone Wants to Do the Model Work, Not
  the Data Work”: Data Cascades in High-Stakes AI}}.
\newblock \bibinfo{publisher}{Association for Computing Machinery},
  \bibinfo{address}{New York, NY, USA}.
\newblock
\showISBNx{9781450380966}
\urldef\tempurl%
\url{https://doi.org/10.1145/3411764.3445518}
\showURL{%
\tempurl}


\bibitem[\protect\citeauthoryear{Sanchez, Caramiaux, Fran\c{c}oise, Bevilacqua,
  and Mackay}{Sanchez et~al\mbox{.}}{2021}]%
        {sanchez10.1145/3449236}
\bibfield{author}{\bibinfo{person}{T\'{e}o Sanchez}, \bibinfo{person}{Baptiste
  Caramiaux}, \bibinfo{person}{Jules Fran\c{c}oise},
  \bibinfo{person}{Fr\'{e}d\'{e}ric Bevilacqua}, {and}
  \bibinfo{person}{Wendy~E. Mackay}.} \bibinfo{year}{2021}\natexlab{}.
\newblock \showarticletitle{How Do People Train a Machine? Strategies and
  (Mis)Understandings}.
\newblock \bibinfo{journal}{\emph{Proc. ACM Hum.-Comput. Interact.}}
  \bibinfo{volume}{5}, \bibinfo{number}{CSCW1}, Article
  \bibinfo{articleno}{162} (\bibinfo{date}{apr} \bibinfo{year}{2021}),
  \bibinfo{numpages}{26}~pages.
\newblock
\urldef\tempurl%
\url{https://doi.org/10.1145/3449236}
\showDOI{\tempurl}


\bibitem[\protect\citeauthoryear{Sanh, Debut, Chaumond, and Wolf}{Sanh
  et~al\mbox{.}}{2020}]%
        {sanh:distilbert:2020}
\bibfield{author}{\bibinfo{person}{Victor Sanh}, \bibinfo{person}{Lysandre
  Debut}, \bibinfo{person}{Julien Chaumond}, {and} \bibinfo{person}{Thomas
  Wolf}.} \bibinfo{year}{2020}\natexlab{}.
\newblock \bibinfo{title}{DistilBERT, a distilled version of BERT: smaller,
  faster, cheaper and lighter}.
\newblock
\newblock
\showeprint[arxiv]{1910.01108}~[cs.CL]


\bibitem[\protect\citeauthoryear{Shen, DeVos, Eslami, and Holstein}{Shen
  et~al\mbox{.}}{2021}]%
        {shen10.1145/3479577}
\bibfield{author}{\bibinfo{person}{Hong Shen}, \bibinfo{person}{Alicia DeVos},
  \bibinfo{person}{Motahhare Eslami}, {and} \bibinfo{person}{Kenneth
  Holstein}.} \bibinfo{year}{2021}\natexlab{}.
\newblock \showarticletitle{Everyday Algorithm Auditing: Understanding the
  Power of Everyday Users in Surfacing Harmful Algorithmic Behaviors}.
\newblock \bibinfo{journal}{\emph{Proc. ACM Hum.-Comput. Interact.}}
  \bibinfo{volume}{5}, \bibinfo{number}{CSCW2}, Article
  \bibinfo{articleno}{433} (\bibinfo{date}{oct} \bibinfo{year}{2021}),
  \bibinfo{numpages}{29}~pages.
\newblock
\urldef\tempurl%
\url{https://doi.org/10.1145/3479577}
\showDOI{\tempurl}


\bibitem[\protect\citeauthoryear{Shneiderman}{Shneiderman}{2020}]%
        {shneiderman2020bridging}
\bibfield{author}{\bibinfo{person}{Ben Shneiderman}.}
  \bibinfo{year}{2020}\natexlab{}.
\newblock \showarticletitle{Bridging the Gap Between Ethics and Practice:
  Guidelines for Reliable, Safe, and Trustworthy Human-Centered AI Systems}.
\newblock \bibinfo{journal}{\emph{ACM Transactions on Interactive Intelligent
  Systems (TiiS)}} \bibinfo{volume}{10}, \bibinfo{number}{4}
  (\bibinfo{year}{2020}), \bibinfo{pages}{1--31}.
\newblock


\bibitem[\protect\citeauthoryear{Socher, Perelygin, Wu, Chuang, Manning, Ng,
  and Potts}{Socher et~al\mbox{.}}{2013}]%
        {socher:sst-2:2013}
\bibfield{author}{\bibinfo{person}{Richard Socher}, \bibinfo{person}{Alex
  Perelygin}, \bibinfo{person}{Jean Wu}, \bibinfo{person}{Jason Chuang},
  \bibinfo{person}{Christopher~D. Manning}, \bibinfo{person}{Andrew Ng}, {and}
  \bibinfo{person}{Christopher Potts}.} \bibinfo{year}{2013}\natexlab{}.
\newblock \showarticletitle{Recursive Deep Models for Semantic Compositionality
  Over a Sentiment Treebank}. In \bibinfo{booktitle}{\emph{Proc. EMNLP'13}}.
  \bibinfo{publisher}{Association for Computational Linguistics},
  \bibinfo{address}{Seattle, Washington, USA}, \bibinfo{pages}{1631--1642}.
\newblock
\urldef\tempurl%
\url{https://aclanthology.org/D13-1170}
\showURL{%
\tempurl}


\bibitem[\protect\citeauthoryear{Sokol and Flach}{Sokol and Flach}{2020}]%
        {sokol10.1145/3351095.3372870}
\bibfield{author}{\bibinfo{person}{Kacper Sokol} {and} \bibinfo{person}{Peter
  Flach}.} \bibinfo{year}{2020}\natexlab{}.
\newblock \showarticletitle{Explainability Fact Sheets: A Framework for
  Systematic Assessment of Explainable Approaches}. In
  \bibinfo{booktitle}{\emph{Proceedings of the 2020 Conference on Fairness,
  Accountability, and Transparency}} (Barcelona, Spain)
  \emph{(\bibinfo{series}{FAT* '20})}. \bibinfo{publisher}{Association for
  Computing Machinery}, \bibinfo{address}{New York, NY, USA},
  \bibinfo{pages}{56–67}.
\newblock
\showISBNx{9781450369367}
\urldef\tempurl%
\url{https://doi.org/10.1145/3351095.3372870}
\showDOI{\tempurl}


\bibitem[\protect\citeauthoryear{Thomas and Uminsky}{Thomas and
  Uminsky}{2020}]%
        {thomas2020problem}
\bibfield{author}{\bibinfo{person}{Rachel Thomas} {and} \bibinfo{person}{David
  Uminsky}.} \bibinfo{year}{2020}\natexlab{}.
\newblock \showarticletitle{The Problem with Metrics is a Fundamental Problem
  for AI}.
\newblock \bibinfo{journal}{\emph{arXiv preprint arXiv:2002.08512}}
  (\bibinfo{year}{2020}).
\newblock


\bibitem[\protect\citeauthoryear{VanderPlas, Granger, Heer, Moritz,
  Wongsuphasawat, Satyanarayan, Lees, Timofeev, Welsh, and Sievert}{VanderPlas
  et~al\mbox{.}}{2018}]%
        {vanderplas:altair:2018}
\bibfield{author}{\bibinfo{person}{Jacob VanderPlas}, \bibinfo{person}{Brian
  Granger}, \bibinfo{person}{Jeffrey Heer}, \bibinfo{person}{Dominik Moritz},
  \bibinfo{person}{Kanit Wongsuphasawat}, \bibinfo{person}{Arvind
  Satyanarayan}, \bibinfo{person}{Eitan Lees}, \bibinfo{person}{Ilia Timofeev},
  \bibinfo{person}{Ben Welsh}, {and} \bibinfo{person}{Scott Sievert}.}
  \bibinfo{year}{2018}\natexlab{}.
\newblock \showarticletitle{Altair: Interactive statistical visualizations for
  python}.
\newblock \bibinfo{journal}{\emph{Journal of open source software}}
  \bibinfo{volume}{3}, \bibinfo{number}{32} (\bibinfo{year}{2018}),
  \bibinfo{pages}{1057}.
\newblock


\bibitem[\protect\citeauthoryear{Viljoen}{Viljoen}{2021}]%
        {viljoen2021relational}
\bibfield{author}{\bibinfo{person}{Salome Viljoen}.}
  \bibinfo{year}{2021}\natexlab{}.
\newblock \showarticletitle{A Relational Theory of Data Governance}.
\newblock \bibinfo{journal}{\emph{The Yale Law Journal}} \bibinfo{volume}{131},
  \bibinfo{number}{573} (\bibinfo{year}{2021}).
\newblock


\bibitem[\protect\citeauthoryear{Wallace, McCarthy, Wright, and
  Olivier}{Wallace et~al\mbox{.}}{2013}]%
        {Wallace:design_probes:2013}
\bibfield{author}{\bibinfo{person}{Jayne Wallace}, \bibinfo{person}{John
  McCarthy}, \bibinfo{person}{Peter~C. Wright}, {and} \bibinfo{person}{Patrick
  Olivier}.} \bibinfo{year}{2013}\natexlab{}.
\newblock \showarticletitle{Making Design Probes Work}. In
  \bibinfo{booktitle}{\emph{Proceedings of the SIGCHI Conference on Human
  Factors in Computing Systems}} (Paris, France) \emph{(\bibinfo{series}{Proc.
  CHI '13})}. \bibinfo{publisher}{Association for Computing Machinery},
  \bibinfo{address}{New York, NY, USA}, \bibinfo{pages}{3441–3450}.
\newblock
\showISBNx{9781450318990}
\urldef\tempurl%
\url{https://doi.org/10.1145/2470654.2466473}
\showDOI{\tempurl}


\bibitem[\protect\citeauthoryear{Wang and Moulden}{Wang and Moulden}{2021}]%
        {wang2021ai}
\bibfield{author}{\bibinfo{person}{Jennifer Wang} {and} \bibinfo{person}{Angela
  Moulden}.} \bibinfo{year}{2021}\natexlab{}.
\newblock \showarticletitle{AI Trust Score: A User-Centered Approach to
  Building, Designing, and Measuring the Success of Intelligent Workplace
  Features}. In \bibinfo{booktitle}{\emph{Extended Abstracts of the 2021 CHI
  Conference on Human Factors in Computing Systems}}. \bibinfo{pages}{1--7}.
\newblock


\bibitem[\protect\citeauthoryear{Whittaker, Crawford, Dobbe, Fried, Kaziunas,
  Mathur, West, Richardson, Schultz, and Schwartz}{Whittaker
  et~al\mbox{.}}{2018}]%
        {whittaker2018ai}
\bibfield{author}{\bibinfo{person}{Meredith Whittaker}, \bibinfo{person}{Kate
  Crawford}, \bibinfo{person}{Roel Dobbe}, \bibinfo{person}{Genevieve Fried},
  \bibinfo{person}{Elizabeth Kaziunas}, \bibinfo{person}{Varoon Mathur},
  \bibinfo{person}{Sarah~Mysers West}, \bibinfo{person}{Rashida Richardson},
  \bibinfo{person}{Jason Schultz}, {and} \bibinfo{person}{Oscar Schwartz}.}
  \bibinfo{year}{2018}\natexlab{}.
\newblock \bibinfo{booktitle}{\emph{AI Now Report 2018}}.
\newblock \bibinfo{publisher}{AI Now Institute at New York University New
  York}.
\newblock


\bibitem[\protect\citeauthoryear{Wolf, Debut, Sanh, Chaumond, Delangue, Moi,
  Cistac, Rault, Louf, Funtowicz, Davison, Shleifer, von Platen, Ma, Jernite,
  Plu, Xu, Scao, Gugger, Drame, Lhoest, and Rush}{Wolf et~al\mbox{.}}{2020}]%
        {wolf:huggingfaces:2020}
\bibfield{author}{\bibinfo{person}{Thomas Wolf}, \bibinfo{person}{Lysandre
  Debut}, \bibinfo{person}{Victor Sanh}, \bibinfo{person}{Julien Chaumond},
  \bibinfo{person}{Clement Delangue}, \bibinfo{person}{Anthony Moi},
  \bibinfo{person}{Pierric Cistac}, \bibinfo{person}{Tim Rault},
  \bibinfo{person}{Rémi Louf}, \bibinfo{person}{Morgan Funtowicz},
  \bibinfo{person}{Joe Davison}, \bibinfo{person}{Sam Shleifer},
  \bibinfo{person}{Patrick von Platen}, \bibinfo{person}{Clara Ma},
  \bibinfo{person}{Yacine Jernite}, \bibinfo{person}{Julien Plu},
  \bibinfo{person}{Canwen Xu}, \bibinfo{person}{Teven~Le Scao},
  \bibinfo{person}{Sylvain Gugger}, \bibinfo{person}{Mariama Drame},
  \bibinfo{person}{Quentin Lhoest}, {and} \bibinfo{person}{Alexander~M. Rush}.}
  \bibinfo{year}{2020}\natexlab{}.
\newblock \bibinfo{title}{HuggingFace's Transformers: State-of-the-art Natural
  Language Processing}.
\newblock
\newblock
\showeprint[arxiv]{1910.03771}~[cs.CL]


\bibitem[\protect\citeauthoryear{Yang, Negoescu, and Ahammad}{Yang
  et~al\mbox{.}}{2021}]%
        {yang2021intellige}
\bibfield{author}{\bibinfo{person}{Jilei Yang}, \bibinfo{person}{Diana
  Negoescu}, {and} \bibinfo{person}{Parvez Ahammad}.}
  \bibinfo{year}{2021}\natexlab{}.
\newblock \showarticletitle{Intellige: A User-Facing Model Explainer for
  Narrative Explanations}.
\newblock \bibinfo{journal}{\emph{arXiv preprint arXiv:2105.12941}}
  (\bibinfo{year}{2021}).
\newblock


\bibitem[\protect\citeauthoryear{Yang, Suh, Chen, and Ramos}{Yang
  et~al\mbox{.}}{2018}]%
        {yang2018grounding}
\bibfield{author}{\bibinfo{person}{Qian Yang}, \bibinfo{person}{Jina Suh},
  \bibinfo{person}{Nan-Chen Chen}, {and} \bibinfo{person}{Gonzalo Ramos}.}
  \bibinfo{year}{2018}\natexlab{}.
\newblock \showarticletitle{Grounding Interactive Machine Learning Tool Design
  in How Non-experts Actually Build Models}. In
  \bibinfo{booktitle}{\emph{Proceedings of the 2018 Designing Interactive
  Systems Conference}}. \bibinfo{pages}{573--584}.
\newblock


\bibitem[\protect\citeauthoryear{Zhang, Albarghouthi, and D'Antoni}{Zhang
  et~al\mbox{.}}{2021}]%
        {zhang2021certified}
\bibfield{author}{\bibinfo{person}{Yuhao Zhang}, \bibinfo{person}{Aws
  Albarghouthi}, {and} \bibinfo{person}{Loris D'Antoni}.}
  \bibinfo{year}{2021}\natexlab{}.
\newblock \showarticletitle{Certified Robustness to Programmable
  Transformations in LSTMs}.
\newblock \bibinfo{journal}{\emph{arXiv preprint arXiv:2102.07818}}
  (\bibinfo{year}{2021}).
\newblock


\bibitem[\protect\citeauthoryear{Zhang, Liao, and Bellamy}{Zhang
  et~al\mbox{.}}{2020}]%
        {zhang10.1145/3351095.3372852}
\bibfield{author}{\bibinfo{person}{Yunfeng Zhang}, \bibinfo{person}{Q.~Vera
  Liao}, {and} \bibinfo{person}{Rachel K.~E. Bellamy}.}
  \bibinfo{year}{2020}\natexlab{}.
\newblock \showarticletitle{Effect of Confidence and Explanation on Accuracy
  and Trust Calibration in AI-Assisted Decision Making}. In
  \bibinfo{booktitle}{\emph{Proceedings of the 2020 Conference on Fairness,
  Accountability, and Transparency}} (Barcelona, Spain)
  \emph{(\bibinfo{series}{FAT* '20})}. \bibinfo{publisher}{Association for
  Computing Machinery}, \bibinfo{address}{New York, NY, USA},
  \bibinfo{pages}{295–305}.
\newblock
\showISBNx{9781450369367}
\urldef\tempurl%
\url{https://doi.org/10.1145/3351095.3372852}
\showDOI{\tempurl}


\end{thebibliography}

\end{document}